\documentclass[aps,prx,twocolumn,showpacs]{revtex4-1}
\pdfoutput=1
\usepackage{hyperref}
\hypersetup{
colorlinks = true, linkcolor = [rgb]{0,0,1}, citecolor = [rgb]{0,0,1},
                            urlcolor = [rgb]{0,0,1}}
\usepackage{graphicx,epsf}
\usepackage{amsmath}
\usepackage{amssymb}
\usepackage{color}
\usepackage{esint}
\usepackage{wasysym}

\newcommand{\bS}{{\bf S}}
\newcommand{\bR}{{\bf R}}
\newcommand{\bq}{{\bf q}}
\newcommand{\bQ}{{\bf Q}}
\newcommand{\bp}{{\bf p}}
\newcommand{\bk}{{\bf k}}

\newcommand{\taub}{\mbox{\boldmath $\tau $}}

\begin{document}

\title{Entanglement entropy for the valence bond solid phases of
         two-dimensional dimerized Heisenberg antiferromagnets}

\author{Leonardo S. G. Leite  and R. L. Doretto}
\affiliation{Instituto de F\'isica Gleb Wataghin,
                  Universidade Estadual de Campinas,
                  13083-859 Campinas, SP, Brazil}

\date{\today}

\begin{abstract}
We calculate the bipartite von Neumann and second R\'enyi entanglement
entropies of the ground states of spin-1/2 dimerized Heisenberg antiferromagnets on a
square lattice.  Two distinct dimerization patterns are considered:
columnar and staggered. In both cases, we concentrate on the valence
bond solid (VBS) phase and describe such a phase with the bond-operator
representation. Within this formalism, the original spin Hamiltonian is
mapped into an effective interacting boson model for the triplet 
excitations. We study the effective Hamiltonian at the
harmonic approximation and determine the spectrum of the elementary
triplet excitations. 
We then follow an analytical procedure, which is based on a modified
spin-wave theory for finite systems and was originally employed to calculate the
entanglement entropies of magnetic ordered phases, and calculate the
entanglement entropies of the VBS ground states. In particular, we consider
one-dimensional (line) subsystems within the square lattice, a choice
that allows us to consider line subsystems with sizes up to $L' = 1000$.  
We combine such a procedure with the results of the  bond-operator
formalism at the harmonic level and show that, for both dimerized
Heisenberg models, the entanglement entropies of the corresponding VBS
ground states obey an area law as expected for gapped phases. 
For both columnar-dimer and staggered-dimer models, we also show that the
entanglement entropies increase but they seem to not diverge as the dimerization
decreases and the system approaches the N\'eel--VBS quantum phase
transition. Finally, the entanglement spectra associated with the VBS ground
states are presented. 
\end{abstract}

\maketitle

\section{Introduction}
\label{sec:intro}

In the last few years, bipartite entanglement entropies have been used
to characterise many-body quantum systems
\cite{rmp08,rmp-area-law,grover13,wen-book,review-nicolas}. 
Such quantities could offer additional information that, in
principle, could not be obtained from correlation functions. In
particular, bipartite entanglement entropies have been employed to 
study interacting spin systems 
\cite{song11,kallin11,lou11,grover11,luitz14,wessel14,luitz15,alet15,melko17}.

The bipartite entanglement entropy for pure states is defined as
follows: Consider, for instance, the ground state $|\Psi\rangle$ of a system $S$,  
a subsystem $A$ (arbitrary size and shape) 
and its complementary $\bar{A}$ such that $S = A \cup \bar{A}$.
The entanglement entropy is defined,
for instance, as the von Neumann entropy \cite{grover13,review-nicolas},   
\begin{equation}
  \mathcal{S} =\mathcal{S}(\rho_A) = - {\rm Tr}\left( \rho_A \ln \rho_A \right),
\label{neumann}  
\end{equation}
where $\rho_A = {\rm Tr}_{\bar{A}}  |\Psi\rangle \langle \Psi |$ is the
reduced density matrix of the subsystem $A$. Alternatively, the
entanglement entropy is defined as the R\'enyi entropy 
\begin{equation}
  \mathcal{S}_\alpha  = \mathcal{S}_\alpha(\rho_A) = \frac{1}{1 - \alpha}  
                    \ln \left[ {\rm Tr}\left(  \rho^\alpha_A  \right) \right],
\label{renyi}  
\end{equation}
where the index $\alpha > 0$ acts as a weight for the probabilities. 
In the limit $\alpha \rightarrow 1$, the R\'enyi entropy \eqref{renyi} 
reduces to the von Neumann entropy \eqref{neumann}.

\begin{figure*}[t]
\centerline{\includegraphics[width=8.5cm]{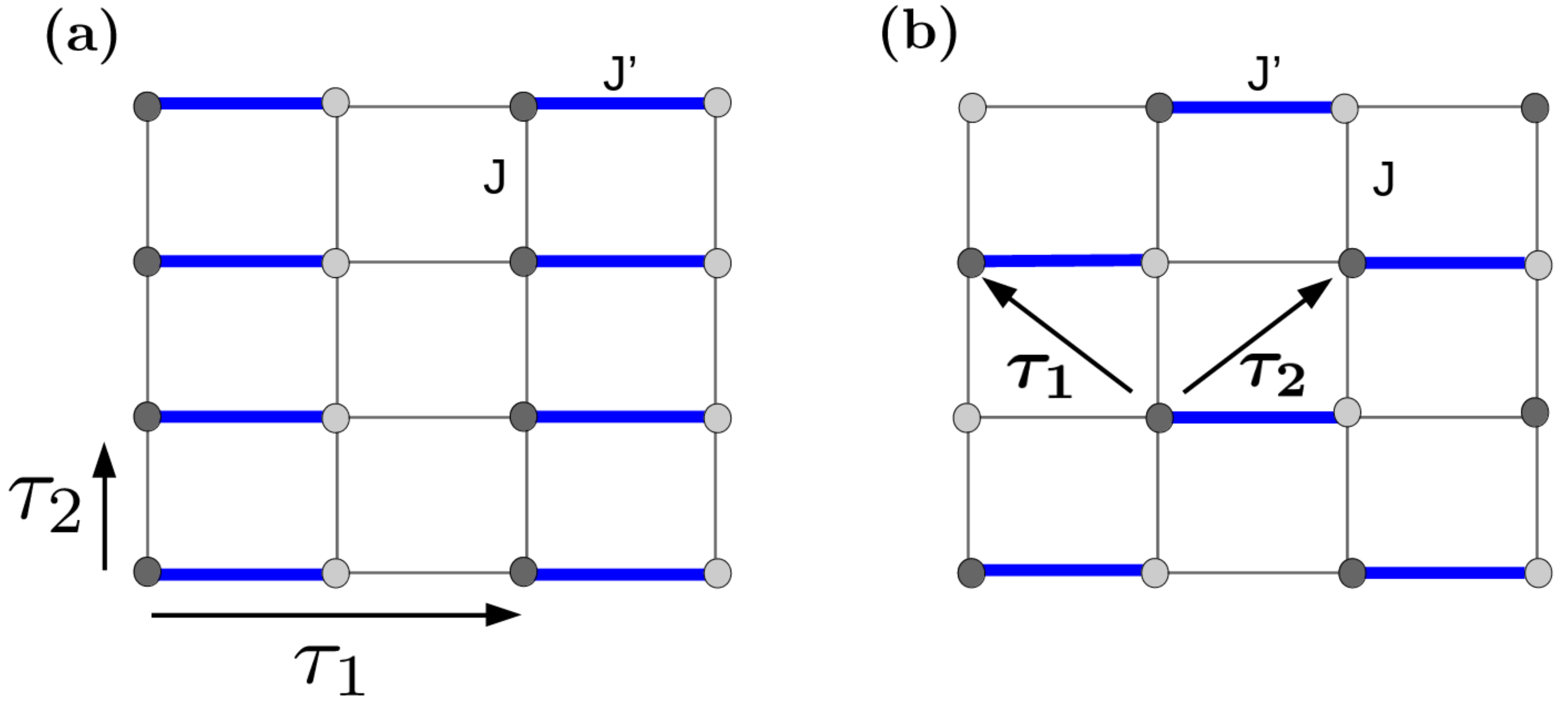} 
                   \hskip1.0cm  
                   \includegraphics[width=8.5cm]{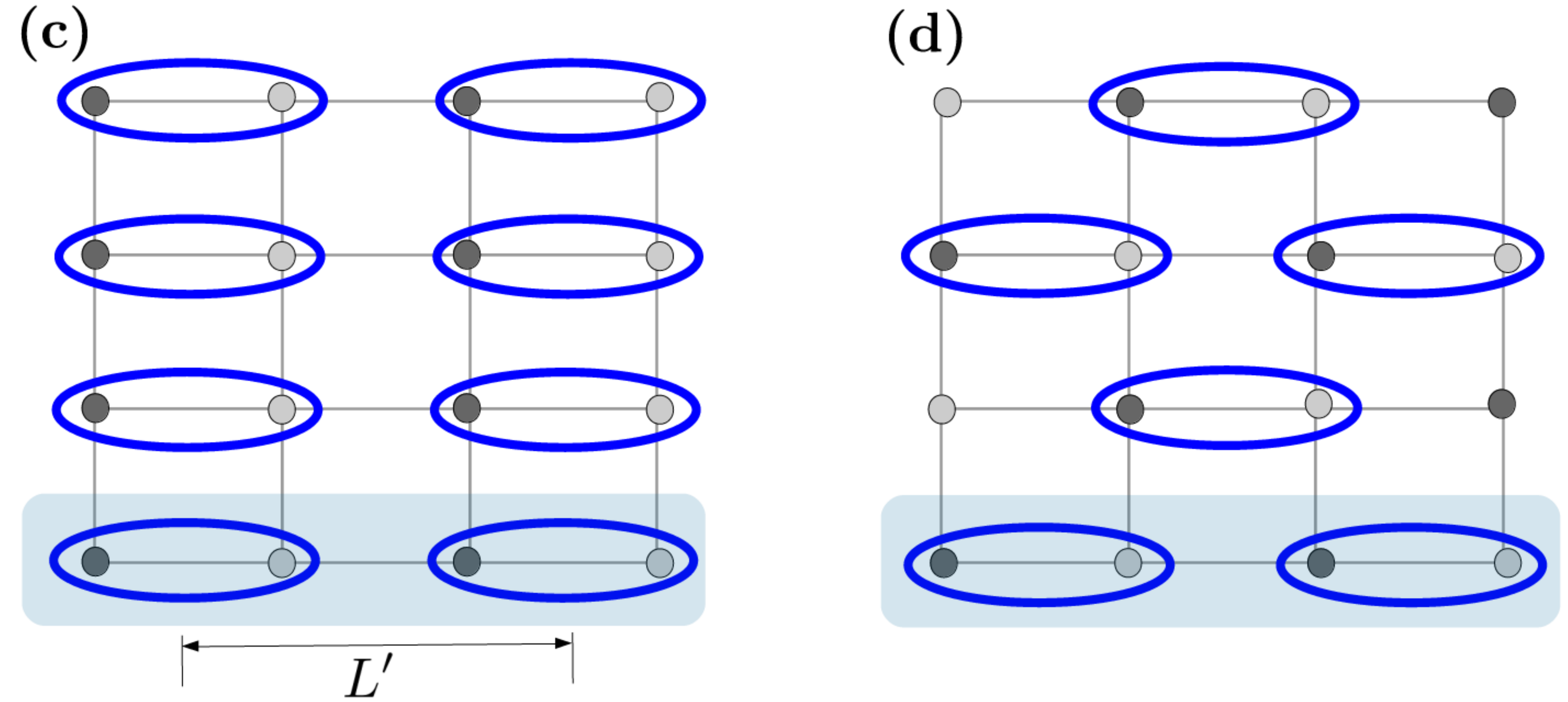}
 } 
\caption{(Color online) Schematic representations of the square lattice
  AFM Heisenberg models with (a) columnar and (b) staggered
  dimerization. The thin (gray) and thick (blue) solid lines
  represent the nearest-neighbor exchange couplings $J = 1$ and $J' >
  J$, respectively. The black and the gray circles respectively
  indicate the spins $\bS^1$ and $\bS^2$ of the underline dimerized
  lattice while $\taub_1$ and $\taub_2$ are the dimer nearest-neighbor
  vectors, see Eqs.~\eqref{tau-col} and \eqref{tau-stag}. 
  Schematic representations of the (c) columnar and (d) staggered
  valence bond solids. The (blue) ellipses represent a singlet stated
  formed by the spins $\bS^1$ and $\bS^2$. The light blue region
  indicates the line subsystem $A$ (one-dimensional dimer chain of length $L'$) 
  considered in the entanglement entropy calculations. 
}
\label{figlattice}
\end{figure*}

An important issue is the scaling of the $T=0$ entanglement entropy 
when the system $S$ is in the thermodynamic limit. 
For instance, for gapped systems in spatial dimensions $d > 1$ 
described by local Hamiltonians, it is found that the entanglement entropy
\eqref{renyi} assumes the general form \cite{wen-book} 
\begin{equation}
 \mathcal{S}_\alpha =  a_\alpha L^{d-1} - \gamma. 
\label{area-law}
\end{equation}
Here the leading term, that depends on the size $L^{d-1}$  of the boundary
between the subsystem $A$ and its complementary $\bar{A}$, 
is the so-called area law \cite{rmp-area-law}.
The coefficient $a_\alpha$ is a non-universal constant. 
The second term $\gamma > 0$ is a universal constant known as the
topological entanglement entropy \cite{kitaev06}. 
Such a quantity indicates whether the ground state has non-trivial
topological order \cite{wen-book}.
An example of a topologically ordered phase (a phase that cannot 
be characterised by a local order parameter) is a 
gapped Z$_2$ spin liquid with $\gamma = \ln 2$ \cite{melko17}.

The entanglement entropy has also been used to characterize gapless
systems. In particular, it was found that the entanglement entropies
of the ground state of the spin-$1/2$ antiferromagnet (AFM) Heisenberg model with
nearest-neighbor  interactions on a square lattice obey an area
law with additive logarithmic corrections \cite{song11,kallin11}.
Later, Metliski and Grover \cite{grover11} analytically calculated the
R\'enyi entanglement entropy \eqref{renyi} of a phase that spontaneously
breaks a continuous symmetry and, for a corner-free subsystem, it was
showed that 
\begin{equation}
 \mathcal{S}_\alpha  =  a_\alpha L^{d-1} 
                        +  \frac{1}{2}n_G\ln\left( \frac{\rho_s}{v}L^{d-1}\right)
                        + \gamma_\alpha^{ord},
\label{s-neel}
\end{equation}
where $\rho_s$ is the spin stiffness, $v$ is the velocity of the $n_G$
Goldstone modes, and $\gamma_\alpha^{ord}$ is a non-universal constant. 
Interestingly, the coefficient of the additional logarithmic
correction to the area law is equal to one-half the number of
Goldstone modes $n_G$.  
Recently, the entanglement entropy of
the N\'eel phase of spin-$1/2$ square lattice Heisenberg AFMs
has been calculated within a modified spin-wave theory for finite systems 
\cite{luitz15,alet15} and the obtained results are in good
agreement with Eq.~\eqref{s-neel}.

In this paper, we calculate the von Neumann \eqref{neumann} and second
($\alpha = 2$) R\'enyi \eqref{renyi} entanglement entropies of the
ground states of spin-$1/2$ dimerized Heisenberg antiferromagnets on a square lattice
focusing on (quantum paramagnet) valence bond solid (VBS) phases. 
We describe the VBS phases within the bond-operator representation \cite{bond-op},
a formalism that allow us to map the original spin Hamiltonian into an effective
interacting boson model for the triplet (triplon) excitations. We then consider such
an effective boson Hamiltonian at the harmonic approximation and determine
the bipartite entanglement entropies via a procedure similar to the modified
spin-wave theory for finite systems \cite{song11,luitz15,alet15}
employed for symmetry broken phases. In particular, we consider 
one-dimensional (line) subsystems of size $L'$ within the square
lattice and analytically calculate the bipartite entanglement entropies.
Such a procedure also allow us to derive the corresponding 
entanglement spectra.

\subsection{Overview of the results}

In the first part of the paper (Secs.~\ref{sec:model}--\ref{sec:harm}),
we study two square lattice dimerized Heisenberg AFMs with columnar
[Fig.~\ref{figlattice}(a)] and staggered [Fig.~\ref{figlattice}(b)]
dimerization patterns. 
We calculate the dispersion relation of the elementary (triplon) excitations
of the VBS phases of the two dimer-models (Fig.~\ref{fig:spectra}) 
within the bond-operator formalism at the (mean-field) harmonic level.
The triplon energy gaps (Fig.~\ref{fig:gaps}) and 
the quantum critical points where the N\'eel--VBS quantum phase
transition (QPT) takes place (Sec.~\ref{sec:self}) are determined.

In the second part of the paper (Sec.~\ref{sec:ee}), we calculate the
bipartite von Neumann \eqref{neumann} and second R\'enyi \eqref{renyi}
entanglement entropies of the VBS ground states of both dimer-models. 
It is shown here that the combination of the bond-operator results at the
harmonic approximation with an approach similar to the one used in  
Refs.~\cite{song11,luitz15,alet15} for magnetic ordered phases 
provides the area law behaviour for the entanglement entropies, a behaviour
expected for gapped phases. This is indeed our main result. 
Importantly, our results are derived for line (chain)
subsystems $A$, a choice that allows us to determine the
entanglement entropies for very large subsystem sizes.
Furthermore, we show that the entanglement entropies
seem to not diverge as the system approaches the N\'eel--VBS
quantum phase transition,  
but only reaches a maximum value (Fig.~\ref{figS1}). Finally, we show that the
corresponding entanglement spectra for the VBS phases are gapped even
when close to the N\'eel--VBS quantum critical point (Fig.~\ref{fig-es}).

\subsection{Outline}

Our paper is organized as follows: 
In Sec.~\ref{sec:model}, we introduce the square lattice columnar and
staggered dimerized Heisenberg AFMs considered in our study. 
In Sec.~\ref{sec:bond}, the bond-operator representation \cite{bond-op} 
for spin operators is briefly summarized and the effective
interacting boson models corresponding to the 
two dimerized Heisenberg antiferromagnets are derived. 
The analysis of the effective boson models within the harmonic
approximation, in particular, the determination of the triplet (triplon)
excitation spectra, is presented in Sec.~\ref{sec:harm}.
In Sec.~\ref{sec:ee}, we briefly review the procedure employed in
Refs.~\cite{song11,luitz15,alet15} for the calculation of the
entanglement entropies of magnetic ordered phases,  
determine the bipartite entanglement entropies for the VBS phases
of the columnar-dimer and staggered-dimer models,
and discuss the corresponding entanglement spectra.
Finally, in Sec.~\ref{sec:summary}, we provide a brief summary of our
main findings.
A short discussion about the classical dimerized Heisenberg AFMs 
and some technical details of the scheme adopted for the calculation
of the entanglement entropies are included in the three Appendices.

\section{Square lattice dimerized antiferromagnets}
\label{sec:model}

Let us consider the dimerized AFM Heisenberg model on a square lattice:
\begin{equation}
	 H = \sum_{ \langle i \,  j \rangle} J_{ij} \bS_i \cdot \bS_j , 
\label{heisenberg}
\end{equation}
where $\bS_i$ is a spin-$1/2$ operator at site $i$ and the
nearest-neighbor exchange couplings $J_{ij} = J > 0$ and $J' > 0$ are
arranged according to the columnar and staggered patterns respectively
shown in Figs.~\ref{figlattice}(a) and (b). Hereafter, we set $J= 1$.

The Hamiltonian \eqref{heisenberg} is an interesting model system to
study quantum phase transitions \cite{vojta-qpt} since its
ground state depends on the (intra-dimer) exchange coupling $J'$: 
for $J' \sim 1$, the ground state has semiclassical N\'eel magnetic
long-range order while, for $J'\gg 1$, a quantum paramagnetic
(disordered) phase
sets in and the ground state is given by a VBS of
short singlets as illustrated in Figs.~\ref{figlattice}(c) and (d). 
The N\'eel--VBS quantum phase transition takes place at the critical
couplings $J'_c = 1.9096(2)$ (columnar) \cite{wenzel09} and 
$J'_c = 2.5196(2)$ (staggered) \cite{wenzel08}. 
According to the quantum-to-classical mapping, this QPT 
should be in the same universality class of the classical Heisenberg
model in $(2+1)$-dimensions, the so-called $O(3)$  universality class
\cite{doretto11}. 
However, quantum Monte Carlo (QMC) results \cite{wenzel08} indicated
that such a scenario applies only to the columnar-dimer model
[Fig.~\ref{figlattice}(a)]:
For the staggered-dimer model [Fig.~\ref{figlattice}(b)], it was found
that the critical exponents showed small deviations from the ones of the
$O(3)$ universality class, an interesting feature that motivated further investigations 
\cite{doretto11,jiang09,jiang12,yasuda13,sandvik18}.
It was then proposed \cite{doretto11} that the critical exponents of
the staggered-dimer model are indeed the ones of the $O(3)$
universality class, but with anomalously large corrections to scaling
related to cubic triplet interactions, see Eq.~\eqref{hk3} below. Such
a scenario was later confirmed by 
QMC calculations \cite{jiang12,yasuda13,sandvik18}.         
This interesting feature of the critical behaviour of the columnar-dimer and
staggered-dimer models found in finite-size QMC simulations is also 
a motivation for our study.   
In the following, we will focus on the region $J' > 1$ of the phase
diagram of the model \eqref{heisenberg}.

In order to describe the VBS phases of the columnar-dimer and
staggered-dimer Heisenberg AFMs, it is useful to rewrite the Hamiltonian
\eqref{heisenberg} in terms of the underline lattices defined by the
strong couplings $J'$:
\begin{equation}
   H = J' \sum_{i \in \mathcal{D}} \: \bS_i^1 \cdot \bS_i^2 
       +   \sum_{\mu \, \nu}\sum_{i \in \mathcal{D}}  \sum_\tau \:
                \bS_i^\mu \cdot \bS_{i+\tau}^\nu. 
\label{HeisenDimer}
\end{equation}
Here $i$ indicates a site of the dimerized lattice $\mathcal{D}$,
which has two spins per unit cell labeled by the 
indices $\mu$ and $\nu = 1,2$, see Figs.~\ref{figlattice}(a)
and (b). The index $\tau = 1,2$ corresponds to the dimer 
nearest-neighbor vectors $\taub_i$: for the columnar-dimer model
[Fig.~\ref{figlattice}(a)], we have 
\begin{equation}
   \taub_1 = 2 a \hat{x},  \;\;\;\;\;\;\;\;\;\;\;
   \taub_2 = a \hat{y}, 			
\label{tau-col}
\end{equation} 	
whereas, for the staggered-dimer model [Fig.~\ref{figlattice}(b)],
\begin{equation}
  \taub_1 = a ( \hat{y}  - \hat{x} ),  \;\;\;\;\;\;\;\;\;\;\;
  \taub_2 = a ( \hat{x}  + \hat{y} ),
\label{tau-stag}
\end{equation}	
with $a$ being the lattice spacing of the {\sl original} square
lattice (in the following we set $a = 1$). In terms of the
nearest-neighbor vectors \eqref{tau-col} and \eqref{tau-stag}, 
the Hamiltonian \eqref{HeisenDimer} can be explicitly written as 
\begin{align}
   H = \:  &J' \sum_{i \in \mathcal{D}}  \bS_i^1 \cdot \bS_i^2 
  \nonumber \\
	       &+ \sum_{i \in \mathcal{D}} \Bigl( \bS_i^1 \cdot \bS_{i + 2}^1 
                  + \bS_i^2 \cdot \bS_{i + 2}^2  
                  + \bS_i^2 \cdot \bS_{i + 1}^1  \Bigr) ,
\label{h1}
\end{align}
for the columnar-dimer model, and  
\begin{align}
   H = \: J' \sum_{i \in \mathcal{D}}  \bS_i^1 \cdot \bS_i^2  
              &+  \sum_{i \in \mathcal{D}} \Bigl( \bS_i^1 \cdot\bS_{i + 1}^2  
 \nonumber \\
	      &+ \bS_i^2 \cdot \bS_{i + 2}^1 + \bS_i^2 \cdot \bS_{i + 2 - 1}^1 \Bigr)
		\label{h2}
\end{align}
for the staggered-dimer model.

\section{Bond-operator representation}
\label{sec:bond}

The VBS phases of the dimerized Heisenberg AFMs 
\eqref{heisenberg} can be described within the bond-operator representation
for spin operators \cite{bond-op}. In the following, we briefly
summarize this formalism.    

We start by considering two spins-$1/2$: $\bS^1$ and $\bS^2$. 
The Hilbert space of the system is made out of a singlet state $|s\rangle$
and three triplet states $|t_\alpha\rangle$:  
\begin{align}
  | s \:\rangle &= \frac{1}{\sqrt{2}} \left( |\uparrow \downarrow \rangle 
                            - |\downarrow \uparrow \rangle \right), 
  \;\;\;\;\;
  |t_x \rangle  =  \frac{1}{\sqrt{2}}\left( |\downarrow \downarrow\rangle 
                          - |\uparrow \uparrow \rangle \right), 
\nonumber \\
  |t_y \rangle &=  \frac{i}{\sqrt{2}}\left( |\uparrow \uparrow \rangle 
                            + |\downarrow \downarrow \rangle \right),
  \;\;\;\;\;
  |t_z \rangle  = \frac{1}{\sqrt{2}} \left( |\uparrow \downarrow\rangle 
                         + |\downarrow \uparrow \rangle \right).
\label{dimer-states}
\end{align}
It is possible to define a set of boson operators $s^\dagger$ and
$t^\dagger_\alpha$ with $\alpha = x$, $y$, $z$ that creates
the states \eqref{dimer-states} out of a fictitious vacuum
$|0\rangle$, namely, 
\begin{equation}
  | s \rangle =  s^{\dagger} |0 \rangle,  
  \;\;\;\;\;\;\;
  | t_\alpha \rangle =  t_\alpha^{\dagger} |0 \rangle, \quad \alpha = x,y,z.
\end{equation}
The unphysical states of the enlarged Hilbert space are removed via
the introduction of the constraint
\begin{equation}
	  s^\dagger s + \sum_\alpha t_\alpha^\dagger t_\alpha  = 1.
 \label{constraint}
\end{equation}
We then calculate the matrix elements of each component of the two
spins operators within the basis $|s \rangle$ and $| t_\alpha \rangle$, 
i.e., we determine $\langle s | S^1_\alpha | s \rangle$, 
$\langle s | S^1_\alpha | t_\beta \rangle$, $\ldots$, and therefore,
based on the obtained results, conclude that the components of the
spin operators $\bS^1$ and $\bS^2$ can be written in terms of the
boson operators $s^\dagger$ and $t^\dagger_\alpha$ as 
\begin{equation}
    S_\alpha^{1,2} = \pm \left( s^{\dagger} t_\alpha + t_\alpha^\dagger s  
                              \mp i \: \epsilon_{\alpha \beta \gamma}  \:  t_\beta^\dagger t_\gamma  \right).
\label{bondOperators}
\end{equation}
Here the indices $\alpha, \beta, \gamma = x,y,z$,
$\epsilon_{\alpha \beta \gamma}$ is the completely antisymmetric
tensor with $\epsilon_{xyz} = 1$, and the summation convention over
repeated indices is implied. 

The bond-operator representation \eqref{bondOperators} can be
generalized to the lattice case, allowing us to express the 
dimerized Heisenberg models \eqref{h1} and \eqref{h2} 
in terms of the boson operators  $s_i^\dagger$ and $t^\dagger_{i \alpha}$.

\subsection{Effective boson models}
\label{sec:boso}

Substituting Eq.~\eqref{bondOperators} generalized to the lattice case
into the Hamiltonian \eqref{h1} of the columnar-dimer model, we find
that the Hamiltonian can be written as 
\begin{equation}
 H = H_0 + H_2 + H_3 + H_4. 
\label{eff-h}
\end{equation}
Here the $H_n$ terms have $n$ triplet operators and are given by 
\begin{widetext}
\begin{align}	
    H_0 =& - \frac{3}{4} J' \sum_i s_i^\dagger s_i , 
\nonumber \\
    H_2 =& \frac{J'}{4} \sum_i t_{i \alpha  }^{\dagger} t_{i \alpha } 
             + \frac{1}{4} \sum_{i, \tau} g_2(\tau) 
                 \left( s_i s_{i+\tau}^\dagger  t_{i \alpha}^ \dagger t_{i+\tau \alpha}  + {\rm H.c.}  
                  + s_i^\dagger s_{i+\tau}^\dagger   t_{i\alpha} t_{i+\tau  \alpha} + {\rm H.c.} \right) ,
\nonumber \\ 
    H_3 =&  \frac{i}{4}    \epsilon_{\alpha \beta \gamma} \sum_{i,\tau }g_3(\tau)\left[  
                  \left( s_i^\dagger   t_{i \alpha} + t_{i \alpha}^\dagger s_i \right )
                           t_{i + \tau \beta}^\dagger  t_{i + \tau  \gamma}   
                           - (i \leftrightarrow i+\tau) \right] ,
\nonumber \\ 
   H_4 =& -\frac{1}{4} \epsilon_{\alpha \beta \gamma} \: \epsilon_{\alpha \beta' \gamma'}
            \sum_{i,\tau} g_4(\tau)t_{i \beta}^\dagger  t_{i+\tau \beta'}^\dagger   t_{i+\tau \gamma'}  t_{i \gamma},
\label{Heffcompleto}
\end{align}
\end{widetext}
with summation convention over repeated indices implied.
The $g_i(\tau)$ functions are defined as
\begin{eqnarray}
  g_2(\tau) &=& 2 \delta_{\tau, 2} - \delta_{\tau, 1},
\nonumber \\
  g_3(\tau) &=& \delta_{\tau,1},
\\
  g_4(\tau) &=& 2 \delta_{\tau, 2} + \delta_{\tau, 1},
\nonumber 
\end{eqnarray}
with the dimer nearest-neighbor vectors $\taub_i$ given by Eq.~\eqref{tau-col}. 
A similar expression is found for the Hamiltonian \eqref{h2} of the
staggered-dimer model, but now the $g_i(\tau)$ functions read
\begin{eqnarray}
  g_2(\tau) &=& (-1)\left( \delta_{\tau, 1} + \delta_{\tau, 2} + 
                                         \delta_{\tau, 2 - 1} \right),
\nonumber \\
  g_3(\tau) &=& \delta_{\tau, 2} - \delta_{\tau, 1} + \delta_{\tau, 2 - 1},
\\
  g_4(\tau) &=&  g_2(\tau), 
\nonumber
\end{eqnarray}
with the $\taub_i$ vectors defined as in Eq.~\eqref{tau-stag}.
One should note that only the last term of the Hamiltonian \eqref{h1} 
contributes to the cubic term $H_3$ whereas, for the staggered-dimer
model, all three nonlocal terms of the Hamiltonian \eqref{h2} provide
a nonvanishing contribution for $H_3$, a feature that can be understood
on symmetry grounds, see Sec.~II.C from Ref~\cite{doretto11}.
Finally, the constraint \eqref{constraint} is taken into account on
average via a Lagrange multiplier $\mu$, i.e., we add the following
term to the Hamiltonian \eqref{eff-h}
\[
   -\mu \sum_i \left( s_i^\dagger s_i + t_{i \alpha}^\dagger t_{i \alpha}  - 1 \right).
\]

Within the bond-operator formalism, the VBS ground states illustrated in
Figs.~\ref{figlattice}(c) and (d) can be viewed as a condensate of the
singlets $s_i$. We then set 
\begin{equation}  
     s_i^\dagger = s_i  = \langle s_i^\dagger \rangle =  
    \langle s_i \rangle \rightarrow  \sqrt{N_0}  
\label{condensate}
\end{equation}
in the Hamiltonian \eqref{eff-h} and arrive at an
effective boson Hamiltonian only in terms of the triplet 
$t_{i\alpha}$ boson operators. 
Both constants $N_0$ and $\mu$ will be self-consistently calculated  
for a fixed value of the exchange coupling $J'$. 
 	
\begin{figure}[t]
\centerline{\includegraphics[width=6.5cm]{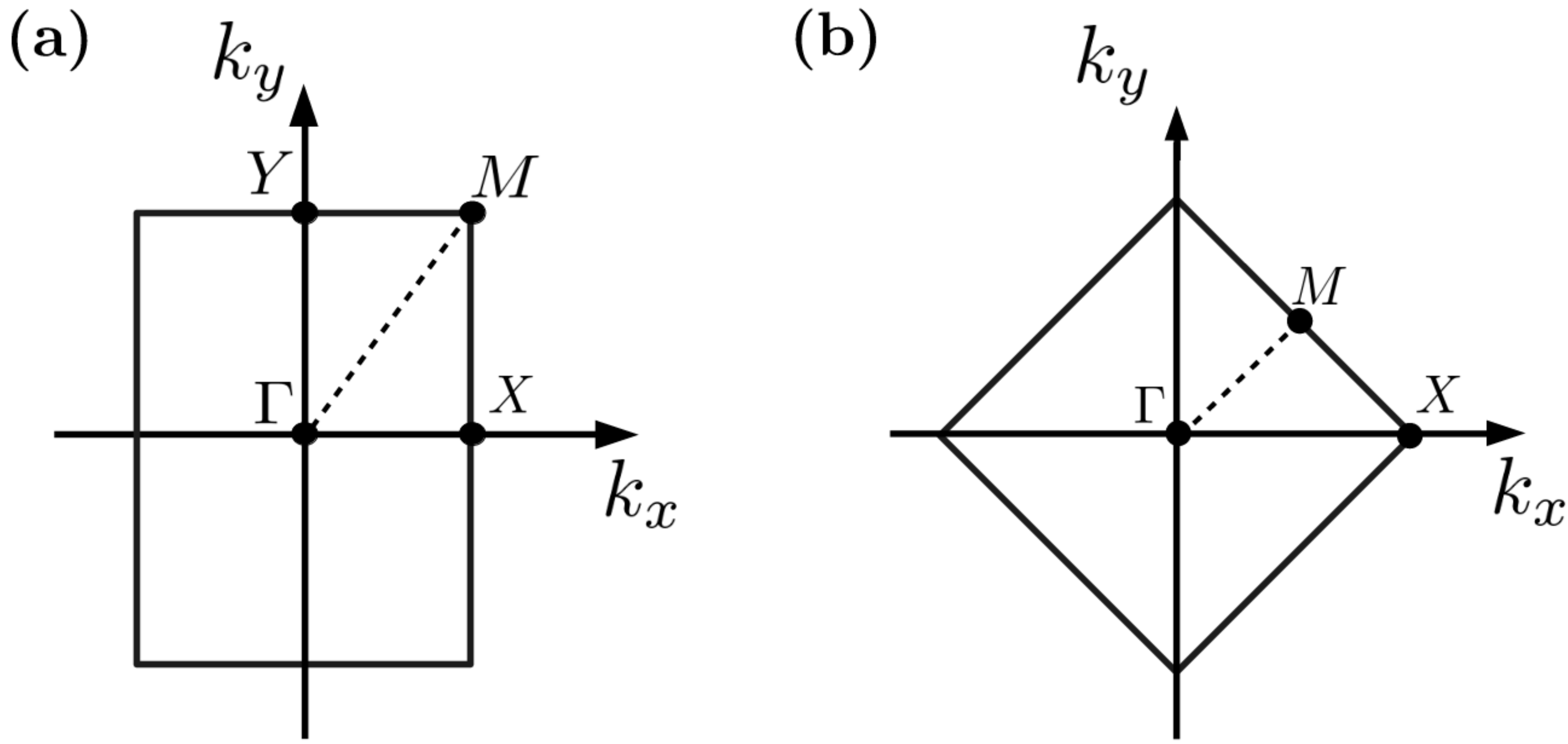}} 
\caption{Schematic representations of the first
  Brillouin zones of the (underline) dimerized lattices for the (a)
  columnar-dimer and (b) staggered-dimer models.
  In panel (a), $\mathbf{X} = (\pi/2,0)$, $\mathbf{M} = (\pi/2,\pi)$,
  and $\mathbf{Y} = (0,\pi)$ while, in panel (b), 
  $\mathbf{X} = (\pi,0)$, $\mathbf{M} = (\pi/2,\pi/2)$.
  The lattice spacing $a$ of the original square lattices is set to 1.}
\label{fig:FBZa}
\end{figure}

Performing a Fourier transform, 
\begin{equation}
   t_{i \alpha}^\dagger = \frac{1}{ \sqrt{N'} } \sum_{\bk \in {\rm BZ}} e^{-i\bk \cdot \bR_i } \: t_{\bk \alpha}^\dagger,
\label{eqOpFourier}
\end{equation}
where $\bR_i$ is a vector of the dimerized lattice, 
$N' = N/2$ is the number of dimers ($N$ is the number of sites of the
{\it original} square lattice), and the momentum sum runs over the
corresponding dimerized first Brillouin zones, see Fig.~\ref{fig:FBZa},
one shows that the four terms \eqref{Heffcompleto}
of the Hamiltonian \eqref{eff-h} assume the form 
\begin{align}
    H_0 = &-\frac{3}{8} J' N_0 N - \mu \frac{N}{2}(N_0 -1),  
\\
    H_2 = &\sum_\bk A_\bk t_{\bk\alpha}^\dagger t_{\bk \alpha} 
              + \frac{1}{2} \sum_\bk B_\bk \:  \left(
             t_{\bk \alpha}^\dagger t_{-\bk \alpha}^\dagger + {\rm H.c.}  \right),
\label{hk2} \\
 H_3 = & \frac{1}{2\sqrt{N'}} \epsilon_{\alpha \beta \gamma}
               \sum_{\bp,\bk} \xi_{\bk - \bp} \left( 
                t_{\bk-\bp \alpha}^\dagger t_{\bp \beta}^\dagger t_{\bk \gamma} + {\rm H.c.} \right), 
\label{hk3} \\
 H_4 = & \frac{1}{2N'} \epsilon_{\alpha \beta \gamma} \epsilon_{\alpha \mu \nu} 
         \sum_{\bp, \bk, \bq} \gamma_\bk \:
         t_{\bp + \bk \beta}^\dagger t_{\bq - \bk \mu}^\dagger t_{\bq \nu} t_{\bp \gamma}.
\label{hk4}  
\end{align}
For the columnar-dimer model, the coefficients $A_\bk$, $B_\bk$,
$\xi_\bk$, and $\gamma_\bk$ are given by 
\begin{align}
    A_\bk &= \frac{J'}{4} + B_\bk - \mu ,  
\nonumber \\
    B_\bk &= \frac{1}{2}  N_0  \left[ 2 \cos k_y - \cos(2k_x) \right], 
\nonumber \\
   \xi_\bk &=   -\sqrt{N_0} \sin(2k_x),
\nonumber \\
   \gamma_\bk &= -\frac{1}{2} \left[ 2 \cos(k_y) + \cos(2k_x) \right], 
\label{bkc}
\end{align}
while, for the staggered-dimer model, we have 
\begin{align}
   A_\bk &= \frac{J'}{4} + B_\bk - \mu ,
\nonumber \\
   B_\bk &= -\frac{1}{2}  N_0  \left[ \cos(2k_x) + \cos(k_x+k_y) + \cos(k_x-k_y)  \right], 
\nonumber \\
   \xi_\bk &=  -\sqrt{N_0} \left[ \sin(2k_x) + \sin(k_x+k_y) + \sin(k_x-k_y)  \right] ,
\nonumber \\
   \gamma_\bk &= -\frac{1}{2} \left[ \cos(2k_x) + \cos(k_x+k_y) + \cos(k_x-k_y)  \right].
\label{bkc2}
\end{align}

It is important to mention that the bond-operator approach to VBS
phases is quite similar to the Holstein--Primakoff one to magnetic
ordered phases, but while the latter considers fluctuations
(spin--waves) above a semiclassical magnetic ordered state, the former
describes excitations above a quantum paramagnetic state. 
Such similarity will be useful in the calculation of the entanglement
entropies below.
Further comparisons between the two approaches can be found in
Sec.~II.B from Ref.~\cite{doretto12}.

The procedure discussed above for the description of a VBS phase
within the bond-operator formalism follows the lines of
Refs.~\cite{bond-op,doretto12}.  
Such a scheme is slightly distinct from the previous bond-operator
study \cite{doretto11} of the dimerized Heisenberg models
\eqref{heisenberg}, which is based on the procedure discussed in Ref.~\cite{kotov98}:
in this case, it is assumed that the boson operators 
$t^\dagger_{i \alpha}$ create triplet excitations out of a 
singlet background $|\Psi_0\rangle = \prod_i s^\dagger_i | 0 \rangle$;
the equivalent of Eq.~\eqref{condensate} reads $s_i^\dagger = s_i  = 1$
and the constraint \eqref{constraint} becomes an inequality, 
$\sum_\alpha t^\dagger_{i \alpha} t_{i \alpha} \le 1$,
which is implemented via an on-site triplet--triplet repulsion term
added to the Hamiltonian.
For both dimer-models at the harmonic approximation, it is found that
the N\'eel--VBS QPT takes place at the critical coupling $J'_c = 3$.
As shown below, the procedure implement in our work provides
better results for $J'_c$ at the (lowest-order) harmonic
approximation.

\section{Harmonic approximation}
\label{sec:harm}

In this section, we consider the effective boson model \eqref{eff-h}
in  the lowest-order approximation, the so-called harmonic
approximation. In this case, we keep the terms of the Hamiltonian
\eqref{eff-h} up to quadratic order in the triplet boson operators
$t_{\bk \alpha}$, namely,  
\begin{equation}
  H \approx H_0 + H_2.
\label{h-harm}
\end{equation}
Since the Hamiltonian \eqref{h-harm} is quadratic in the triplet
operators $t_{\bk \alpha}$, it can be diagonalized via a Bogoliubov
transformation  
\begin{eqnarray}
   b_{\bk \alpha} &=&  u_\bk t_{\bk \alpha} - v_\bk  t_{-\bk \alpha}^\dagger,
\nonumber \\
   b_{\bk \alpha}^\dagger &=&  u_\bk t_{\bk \alpha}^\dagger - v_\bk t_{-\bk \alpha}.
\label{bogoliubovt} 
\end{eqnarray}	
It is then easy to show that the Hamiltonian \eqref{h-harm} assumes
the form
\begin{equation}
   H = \tilde{E}_0 + \sum_{\bk \alpha} \Omega_\bk  b_{\bk \alpha}^\dagger b_{\bk \alpha},
\label{Hdiagonal}
\end{equation}
where 
\begin{equation}
  \tilde{E}_0 = - \frac{3}{8} J' N_0 N - \mu \frac{N}{2}(N_0 - 1) 
                  + \frac{3}{2} \sum_\bk( \Omega_\bk - A_\bk)
\label{en-gs}
\end{equation}
is the ground state energy,
\begin{equation}
  \Omega_\bk = \sqrt{ A_\bk^2 - B_\bk^2 }
\label{Omega}
\end{equation}
is the energy of the triplet (triplon) excitations above
the VBS ground state, and the coefficients $u_\bk$ and $v_\bk$ 
of the Bogoliubov transformation \eqref{bogoliubovt} are given by
\begin{equation}
    u_\bk^2 , v_\bk^2 =  \frac{1}{2} \left( \frac{ A_\bk }{ \Omega_\bk } \pm 1  \right) 			
    \quad  {\rm and} \quad
   u_\bk v_\bk =  \frac{1}{2}\frac{ B_\bk }{ \Omega_\bk }.
\label{coeficients2}	
\end{equation}

Finally, we would like to quote some triplet-triplet ground state
expectation values that will be useful in the determination of the
entanglement entropy, see Sec.~\ref{sec:ee-vbs} below.  With the aid of
Eq.~\eqref{bogoliubovt} and using the fact that  
the ground state of the Hamiltonian \eqref{Hdiagonal} is the vacuum
for the boson operators $b$, one easily shows that  
\begin{equation}
    \langle t_{\bk \alpha }^\dagger t_{\bk \alpha}\rangle = v_\bk^2
    \quad {\rm and} \quad
    \langle t_{\bk \alpha } t_{-\bk \alpha } \rangle =  - u_\bk v_\bk.
\label{tdt} 
\end{equation}

\begin{figure}[t]
 \centerline{\includegraphics[width=7.5cm]{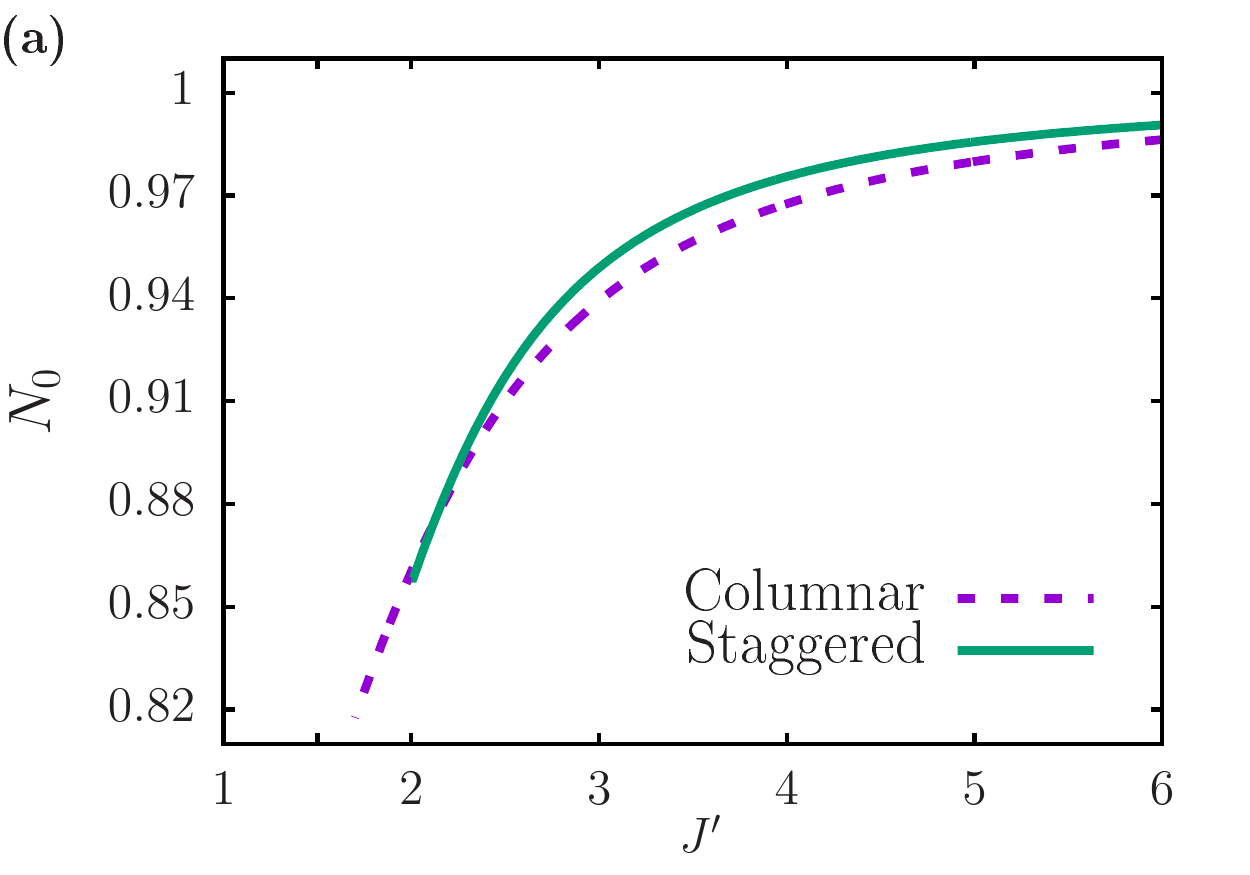}}
 \vskip0.4cm
 \centerline{\includegraphics[width=7.5cm]{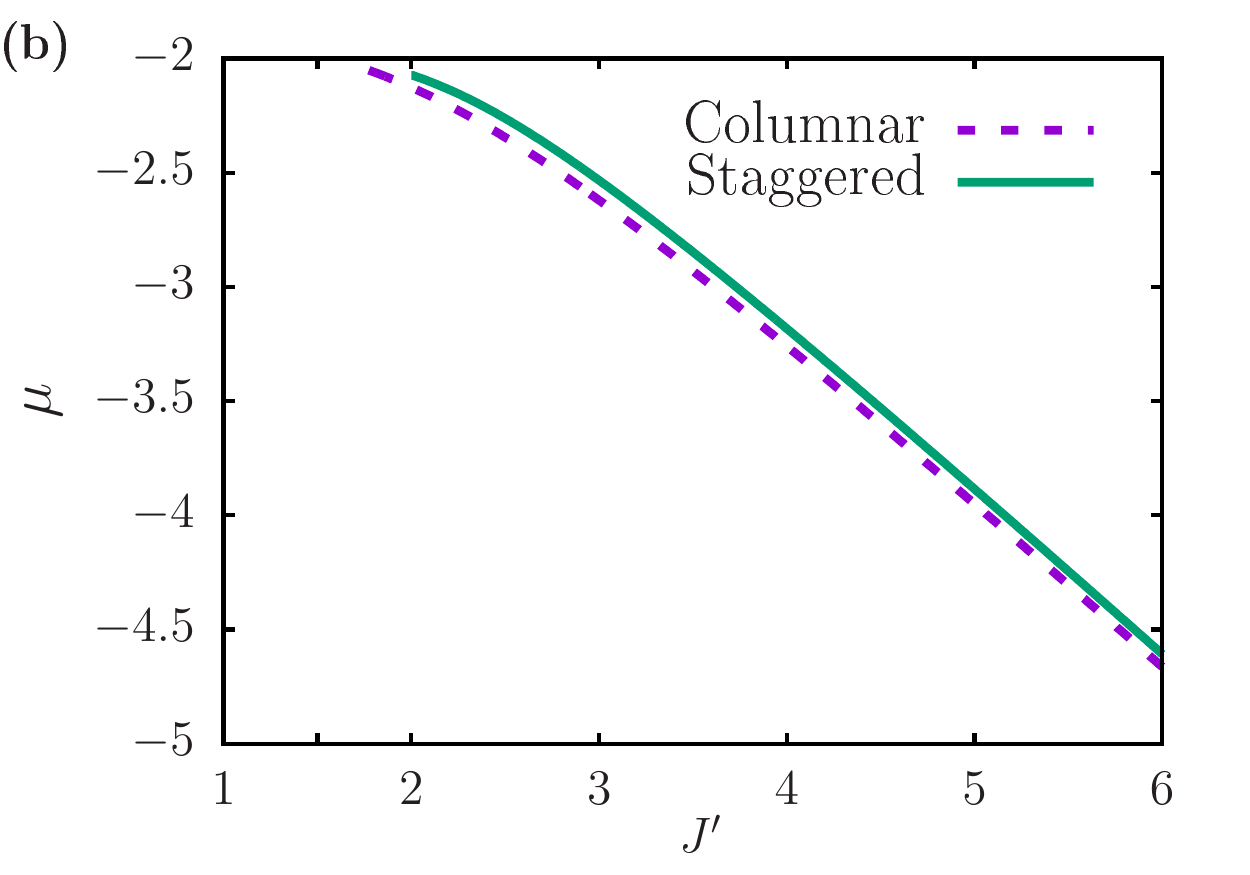}} 
\caption{(Color online) The parameters (a) $N_0$ and (b) $\mu$ as a
  function of the exchange coupling $J'$ determined from the solutions
  of the system of self-consistent equations \eqref{autoconsistencia} 
  within the harmonic approximation. 
  The dashed (magenta) and solid (green) lines respectively correspond
  to the columnar-dimer and staggered-dimer models.} 
\label{fignomu}
\end{figure}

\begin{figure}[t]
\centerline{\includegraphics[width=8.0cm]{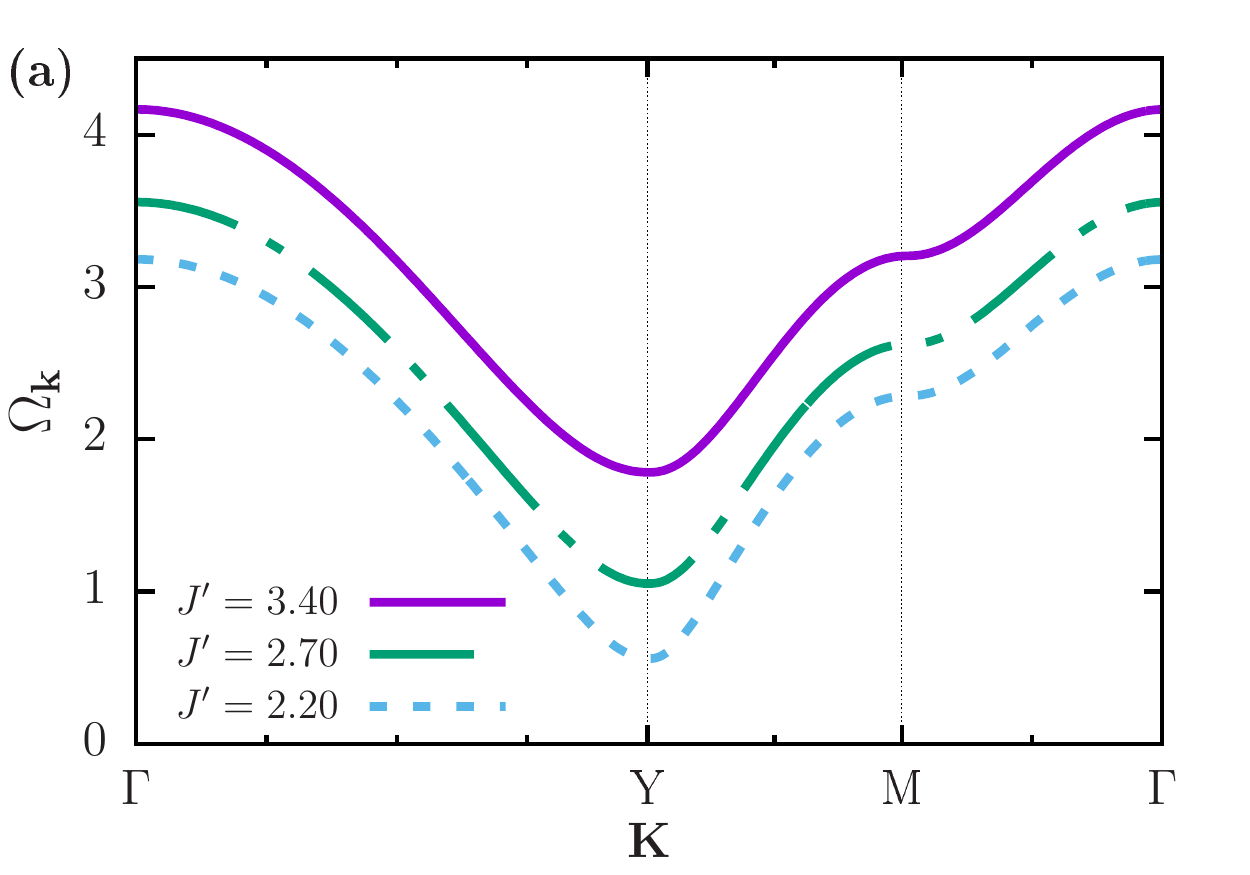}} 
\vskip0.2cm
\centerline{\includegraphics[width=8.0cm]{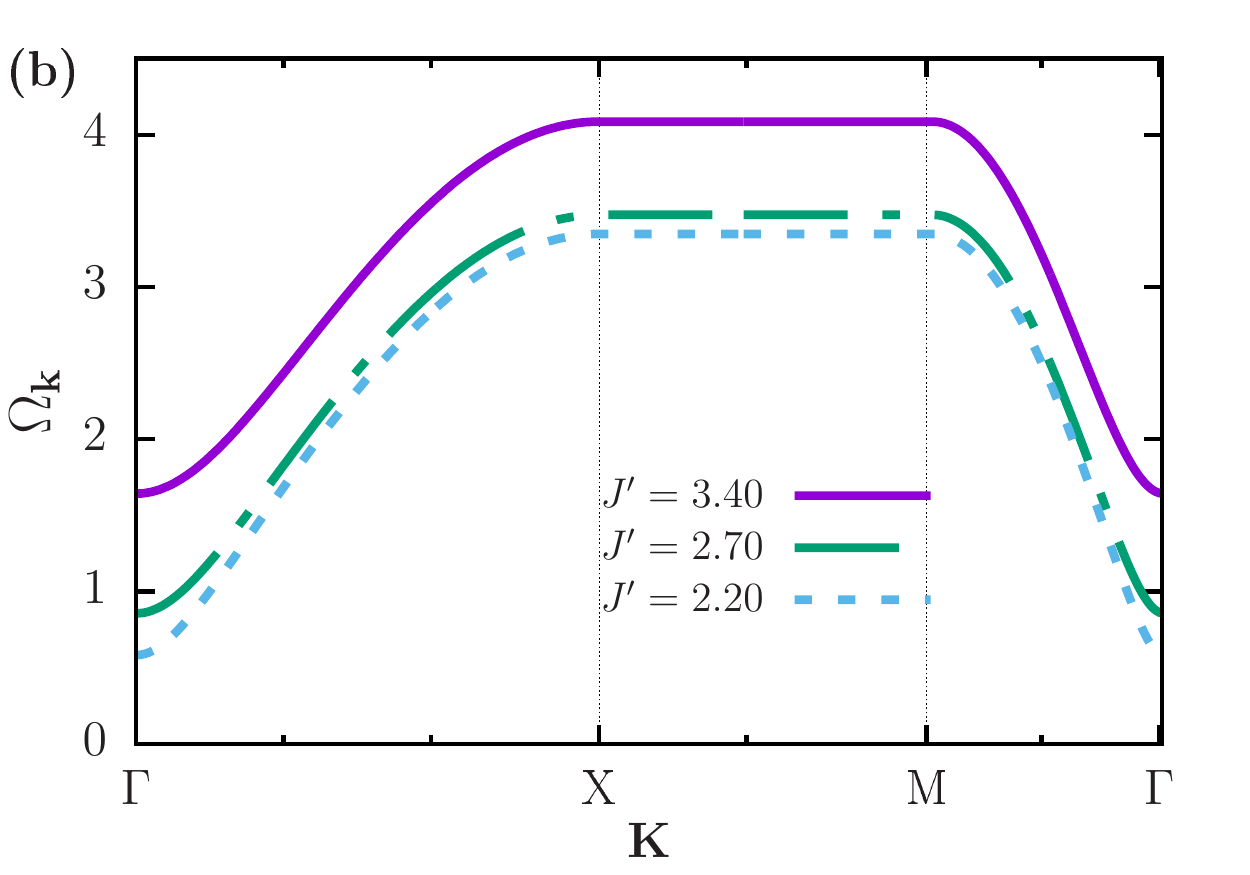}} 
\caption{(Color online) Triplon dispersion relations $\Omega_\bk$
  [Eq.~\eqref{Omega}] along paths in the dimerized first Brillouin zone
  [Fig.~\ref{fig:FBZa}] at the harmonic level for the (a)
  columnar-dimer and (b) staggered-dimer models. 
  Results for three different values of the
  exchange coupling $J'$ are shown: 
  $J'= 2.20$ (dashed blue line),
  $J'= 2.70$ (dotted-dashed green line), and 
  $J'= 3.40$ (solid magenta line).} 
\label{fig:spectra}
\end{figure}

\subsection{Self-consistent equations}
\label{sec:self}

The ground state energy \eqref{en-gs} and the triplon excitation
spectrum \eqref{Omega} are fully determined once we calculate 
the constants $N_0$ and $\mu$ for a fixed value of the exchange
coupling $J'$. 
By minimizing  the ground state energy \eqref{en-gs} with respect to
$\mu$ and $N_0$, we find a system of self-consistent equations,  
\begin{align}
  \mu &= -\frac{3 J'}{4} + \frac{3}{N N_0} \sum_\bk 
         \left[ \frac{B_\bk}{\Omega_\bk}  \left( A_\bk - B_\bk - \Omega_\bk  \right) \right], 
   \nonumber \\
   N_0 &= \frac{3}{N} \sum_\bk \left( 1 - \frac{A_\bk}{\Omega_\bk} \right) +1, 
\label{autoconsistencia}
\end{align}
which are numerically solved. 

The numerical solutions of Eq.~\eqref{autoconsistencia},
i.e., the behaviour of $N_0$ and $\mu$ in terms of $J'$, are
respectively shown in Figs.~\ref{fignomu}(a) and (b).  
As expected, see discussion in Sec.~\ref{sec:model}, 
we find solutions for the system of self-consistent equations
\eqref{autoconsistencia} only for 
$J' \ge 1.70$ (columnar-dimer) and 
$J' \ge 2.00$ (staggered-dimer). 
One sees that     
(i) $N_0 \rightarrow 1$ as the coupling $J'$ increases 
(system deep in the VBS phase)
and (ii) $N_0$ decreases as $J'$ approaches the N\'eel--VBS QPT.

Figure \ref{fig:spectra} shows the triplon excitation spectra
\eqref{Omega} of the columnar-dimer [Fig.~\ref{fig:spectra}(a)] and
the staggered-dimer [Fig.~\ref{fig:spectra}(b)] models
for three different values of $J'$. One sees that
the triplon excitation spectrum is gapped for both models and that the
triplon energy gap $\Delta$ decreases as $J'$ approaches the 
N\'eel--VBS QPT (see details below).
For the columnar-dimer model, the triplon gap $\Delta$ is located
at the $Y$ point [see Fig.~\ref{fig:FBZa}(a)] while, for the staggered-dimer
model, the triplon gap is located at the centre of the first Brillouin zone,
the $\Gamma$ point [see Fig.~\ref{fig:FBZa}(b)]. For both
dimer-models, it is possible to
show that the momentum associated with the triplon gap $\Delta$ is
equal to the ordering wave vector $\bQ$ 
of the corresponding N\'eel magnetic long-range ordered phase that
sets in for $J' < J'_c$, see Appendix~\ref{ap:classical} for details.

\begin{figure}[t]
 \centerline{\includegraphics[width=7.5cm]{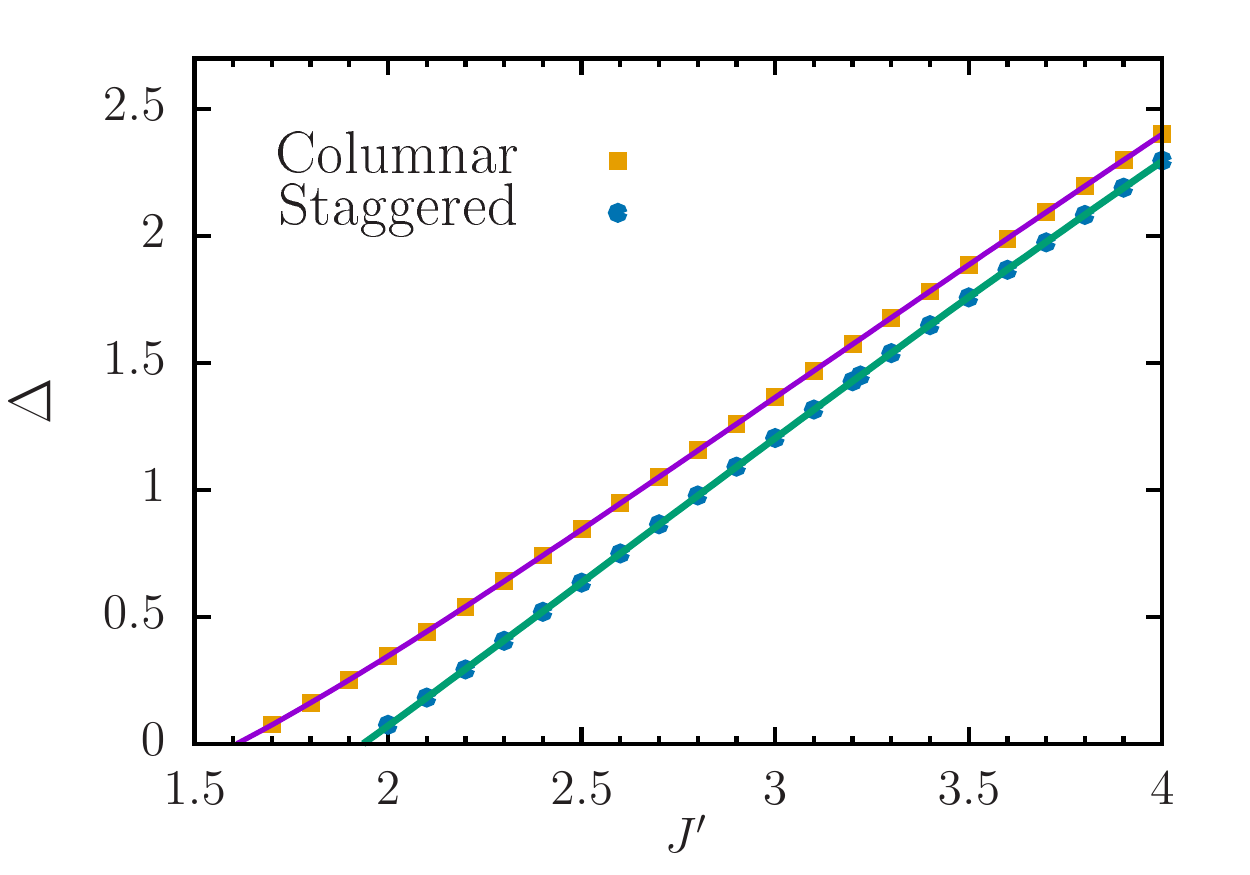}}  
\caption{(Color online) Triplon gaps $\Delta$ as a function of the
  exchange coupling $J'$ for the columnar-dimer (orange squares) and
  staggered-dimer (blue circles) models within the harmonic
  approximation. The solid lines indicate the fits with the expression
 \eqref{fit-gap}.} 
\label{fig:gaps}
\end{figure}

The behaviour of the triplon gaps $\Delta$ as a function of  $J'$ are
displayed in Fig.~\ref{fig:gaps}.  Again, one notices that the triplon
gaps $\Delta$ close as the systems reach the N\'eel--VBS quantum 
critical points. In order to estimate the critical coupling
$J'_c$, we follow the lines of Ref.~\cite{doretto12}, i.e., we assume
a continuous N\'eel--VBS QPT, fit the data with the curve 
\begin{equation}
 \Delta = a_0 + a_1J' + a_2(J')^2 + a_3\frac{1}{J'}, 
\label{fit-gap}
\end{equation}  
and then consider the condition $\Delta = 0$.
Following such a procedure, 
within the harmonic approximation, the critical couplings are 
$J'_c = 1.61$ (columnar) and $J'_c = 1.93$ (staggered),
which are in quite reasonable agreement with the ones determined via
QMC calculations, namely,  
$J'_c = 1.9096(2)$ (columnar) \cite{wenzel09} and 
$J'_c = 2.5196(2)$ (staggered) \cite{wenzel08}. 
Such an agreement is expected due to the small number 
of triplets $t$ in the VBS ground state, $1 - N_0$,
a quantity that could be taken as a control parameter within the
bond-operator formalism.
More accurate results for $J'_c$ can be obtained
within the bond-operator formalism by perturbatively including the
cubic \eqref{hk3} and quartic \eqref{hk4} terms as done, e.g., in
Ref.~\cite{doretto12}.  
Finally, one should also mention that the critical couplings found here are in
better agreement with the QMC simulations than the ones ($J_c' = 3$)
obtained in the previous bond-operator study \cite{doretto11} at the same
approximation level.

\section{Entanglement entropies}
\label{sec:ee}

In this section, we calculate the von Neumann \eqref{neumann} and the
second ($\alpha = 2$) R\'enyi \eqref{renyi} entanglement entropies for
the VBS ground states of both columnar-dimer and staggered-dimer models. 
Both quantities are good measures of entanglement, but the second
R\'enyi entanglement entropy is easier to numerically determined \cite{grover13}.   
For instance, Helmes and Wessel calculated the second R\'enyi
entanglement entropy of a two-dimensional bilayer Heisenberg AFM
\cite{wessel14} based on a QMC procedure introduced in
Ref.~\cite{humeniuk12}. Since we would like to compare our analytical results
with future numerical ones, it is interesting to determined both the   
the von Neumann \eqref{neumann} and the
second ($\alpha = 2$) R\'enyi \eqref{renyi} entanglement entropies
within our scheme.

Due to the similarities between the description of VBS phases within
the bond-operator formalism and the description of magnetic ordered phases within
spin--wave theory (see Sec.~\ref{sec:boso}), we follow the lines of
Refs.~\cite{song11, luitz15,alet15},  
where bipartite entanglement entropies for the N\'eel phase of
two-dimensional Heisenberg AFMs are determined via a
modified spin--wave theory for finite systems.
For completeness, in the following we briefly outline such a scheme
which is indeed based on
Refs.~\cite{peschel01,peschel03,peschel09,barthel06,frerot15}.   

Let us consider a $d$-dimensional system $S$ described by a generic
quadratic Hamiltonian \cite{frerot15}  
\begin{equation}
   H = \sum_{n,m}^N \left[ a_n^\dagger A_{nm} a_m 
         + \frac{1}{2} \left(  a_n^\dagger B_{nm} a_m^\dagger + {\rm H.c.} \right) \right],
\label{generalH1}
\end{equation}
where $a_n$ is a boson operator associated with the site $n$ of a
$d$-dimensional hypercubic lattice with $N$ sites and
$A_{nm}$ and $B_{nm}$ are $N \times N$ matrices. 
We divide the system $S$ into a subsystem $A$ with $N_A < N$ sites and
its complementary $\bar{A}$ such that $S = A \cup \bar{A}$, see Sec.~\ref{sec:intro}.
It is possible to show, e.g., with the aid of coherent states \cite{barthel06,peschel01},  
that the reduced density matrix $\rho_A$ of the subsystem $A$ assumes the form  
\cite{peschel01,peschel03,peschel09,barthel06,frerot15}
\begin{equation}
   \rho_{A} = \mathcal{K} e^{-\mathcal{H}_E}, 
\label{rhoDiagonal}
\end{equation}
where $\mathcal{K}$ is a normalization constant and $\mathcal{H}_E$ is
the so-called {\it entanglement Hamiltonian} \cite{frerot15} 
\begin{equation}
   \mathcal{H}_E = \sum_{i,j}^{N_A} \left[ a_i^\dagger \mathcal{A}_{ij} a_j 
         + \frac{1}{2} \left(  a_i^\dagger \mathcal{B}_{ij} a_j^\dagger + {\rm H.c.} \right) \right],
\label{generalH2}
\end{equation}
with $\mathcal{A}_{ij}$ and $\mathcal{B}_{ij}$ being $N_A \times N_A$ matrices.
Notice that both Hamiltonians $H$ and $\mathcal{H}_E$ have the same
quadratic form and the latter is restricted to the sites $i$ and $j$
of the subsystem $A$. 
Due to this similarity, the Hamiltonians $H$ and $\mathcal{H}_E$ can
then be diagonalized by the same Bogoliubov transformation. In
particular, we have    
\begin{equation}
  \mathcal{H}_E = \sum_k \epsilon_k  b_k^\dagger b_k,
\label{ent-h}
\end{equation}
where the energies $\epsilon_k$ define the 
{\it entanglement spectrum} \cite{review-nicolas,haldane-es}
and the boson operators $a_i$ and $b_k$ are related by a Bogoliubov
transformation, see Eq.~\eqref{Atransformation}. 
Since the reduced density matrix $\rho_A$ has the form \eqref{rhoDiagonal},   
the von Neumann entanglement entropy \eqref{neumann} is simply given
by the expression of the thermal entropy, i.e.,   
\begin{equation}
 \mathcal{S} = \sum_k ( n_k + 1)\ln ( n_k + 1) - n_k\ln n_k,
\label{thermal-s}
\end{equation}
where $n_k = 1/[ \exp(\epsilon_k) - 1]$ is the occupation of the 
$k$ mode. Therefore, once the entanglement spectrum $\epsilon_k$ is
known, the bipartite von Neumann entanglement entropy 
\eqref{neumann} is determined. 
Similar considerations hold for fermionic systems described by
Hamiltonians of the form \eqref{generalH1}
\cite{barthel06,peschel01,cheong04}. 
For a more general expression for the entanglement Hamiltonian,
we refer the reader, e.g., to Ref.~\cite{calabrese18}, where
an approximate entanglement Hamiltonian for a
general lattice model is derived based on a lattice version of the
so-called Bisognano-Wichmann theorem.

Instead of performing the partial trace described above to calculate
the matrix elements $\mathcal{A}_{ij}$ and $\mathcal{B}_{ij}$ and then
find the entanglement entropy \eqref{thermal-s},
we can alternatively determine $\mathcal{S}$
from single-particle Green's functions associated with
the Hamiltonian \eqref{generalH1} 
\cite{barthel06,peschel01,peschel03,peschel09}. 
Indeed, the entanglement spectrum $\epsilon_k$ is related to 
the eigenvalues of the so-called correlation matrix $C$, which is
defined as \cite{barthel06,peschel01,peschel03,peschel09} 
\begin{equation}
  C_{ij} = 4\sum_{s \in  A} \left( f_{is} + g_{is} \right) \left(f_{sj} - g_{sj} \right).
\label{correlationMatrix}
\end{equation}
Here $i$, $j$, and $s$ refer to sites of the subsystem $A$
and $f_{ij}$ and $g_{ij}$ are single-particle Green's functions, 
\begin{equation}
         f_{ij} = \langle a_i^\dagger a_j \rangle + \frac{1}{2}\delta_{ij} 
\quad\quad {\rm and} \quad\quad 
        g_{ij} = \langle a_i  a_j  \rangle,
\label{greenf} 
\end{equation}
As discussed in details in Appendix \ref{ap:pseudoenergy},
one shows that the $N_A$ eigenvalues $\mu^2_k$ of the correlation
matrix $C$ can be written in terms of the entanglement spectrum
$\epsilon_k$ as   
\begin{equation}
 \mu_k =  \coth \left( \frac{\epsilon_k}{2} \right)
     \quad \textrm{or} \quad
  \epsilon_k = \ln \left(\frac{\mu_k +1}{\mu_k -1} \right). 
\label{ekck}
\end{equation}
Substituting Eq.~\eqref{ekck} into the expression \eqref{thermal-s},
one shows that the von Neumann entanglement entropy
\eqref{neumann} reads
\begin{equation}
  \mathcal{S} = \sum_{k = 1}^{N_A}  \sum_{\epsilon = \pm 1} 
        \epsilon\left( \frac{\mu_k + \epsilon }{2} \right)\ln\left( \frac{\mu_k + \epsilon }{2} \right).
\label{Smu-neumann}
\end{equation}
Similarly, one finds that the R\'enyi entanglement entropies \eqref{renyi} 
assume the form \cite{song11, luitz15,alet15}
\begin{equation}
    \mathcal{S}_\alpha = \frac{1}{\alpha-1} \sum_{k = 1}^{N_A} \ln \left[  
                      \left( \frac{\mu_k +1 }{2} \right)^\alpha 
                   - \left( \frac{\mu_k -1 }{2} \right)^\alpha \right].
\label{Sqmu}
\end{equation}
Therefore, the bipartite entanglement entropies are completely
determined, once the eigenvalues $\mu^2_k$ of the correlation matrix
$C$ are known.

Within linear spin--wave theory, the effective boson model that
describes the N\'eel phase of an Heisenberg AFM has the same form as
the Hamiltonian \eqref{generalH1}. 
Due to such similarity, the procedure discussed above was employed to
calculated bipartite entanglement entropies of magnetic ordered phases
of Heisenberg AFMs \cite{song11,luitz15,alet15}.      
In particular, the single--particle Green's functions \eqref{greenf}
can be easily calculated within linear spin--wave theory.
Notice that the same considerations hold for the description of the VBS
phases of the columnar-dimer and staggered-dimer models within the
bond-operator formalism at the harmonic approximation, see
Eq.~\eqref{h-harm}. Therefore, in the next section, we apply the
scheme described above for the VBS phases of the dimerized Heisenberg
models \eqref{h1} and \eqref{h2}.

\subsection{Entanglement entropies of the VBS phases}
\label{sec:ee-vbs}

To determine the bipartite entanglement entropies
for the VBS phases of the columnar-dimer and the staggered-dimer
models, we considerer a line subsystem $A$, i.e.,   
an one-dimensional spin chain with size $L' = 2(N_A - 1)$, 
as illustrated in Fig.~\ref{figlattice}(c).  
Such a partition is quite interesting, since it
allows us to reach very large system sizes \cite{luitz15}   
in addition to analytically determine the entanglement entropies
\cite{alet15}. Indeed, a line subsystem has been employed to study
interacting spin systems \cite{luitz14, luitz15,alet15} and,
in particular, it provides \cite{luitz15}  a prefactor for the
logarithmic term in Eq.~\eqref{s-neel} in good agreement with the
analytical results of Metliski and Grover \cite{grover11}.

The matrix elements \eqref{correlationMatrix} of the correlation matrix $C$
are easily calculated. From Eqs.~\eqref{tdt} and \eqref{greenf}, one shows that 
\begin{align}
	f_{ij} &= + \frac{1}{2N'} \sum_{\bk\in {\rm BZ}} \cos\left[
                 \bk\cdot\left(\bR_i - \bR_j \right) \right] \frac{A_\bk}{\Omega_\bk}, 
\nonumber	\\
	g_{ij} &= -\frac{1}{2N'} \sum_{\bk\in {\rm BZ}} \cos\left[
                 \bk\cdot\left(\bR_i - \bR_j \right) \right] \frac{B_\bk}{\Omega_\bk}, 
\label{transf-f-g} 
\end{align}   
where $\bR_i = 2i\hat{x}$, with $i = 1,2,\ldots,N_A$, is a vector of
the (underline) dimerized lattice of the line subsystem $A$, 
the coefficients $A_\bk$ and $B_\bk$ are given by Eqs.~\eqref{bkc} and
\eqref{bkc2} respectively for the columnar-dimer and 
staggered-dimer models, 
$\Omega_\bk$ is the triplon excitation energy \eqref{Omega}, and the
momentum sums run over the corresponding first Brillouin zones.
Notice that the correlation matrix $C$ is completely determined by the
coefficients $A_\bk$ and $B_\bk$ of the  effective boson model
\eqref{h-harm}.

\begin{figure}[t]
\centerline{\includegraphics[width=7.5cm]{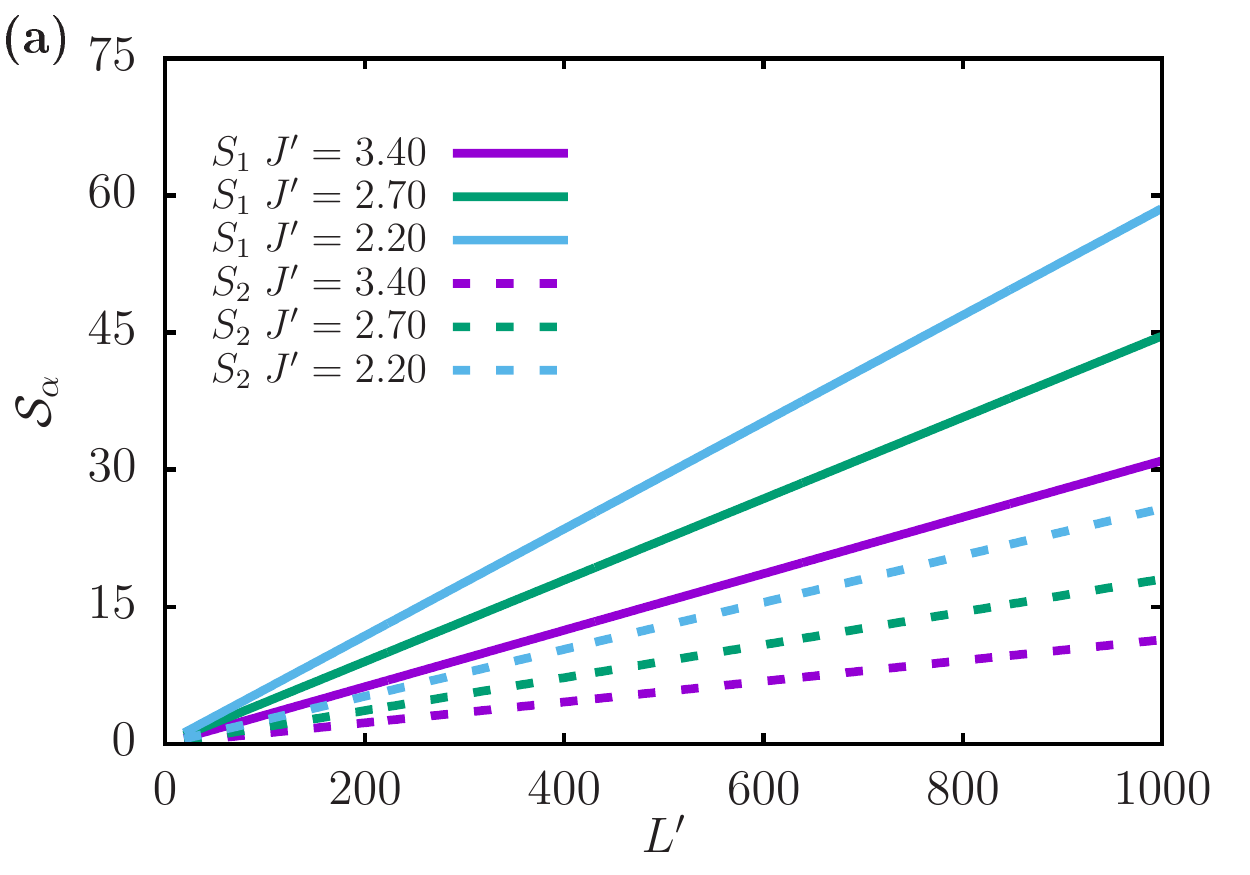}} 
\vskip0.4cm
\centerline{\includegraphics[width=7.5cm]{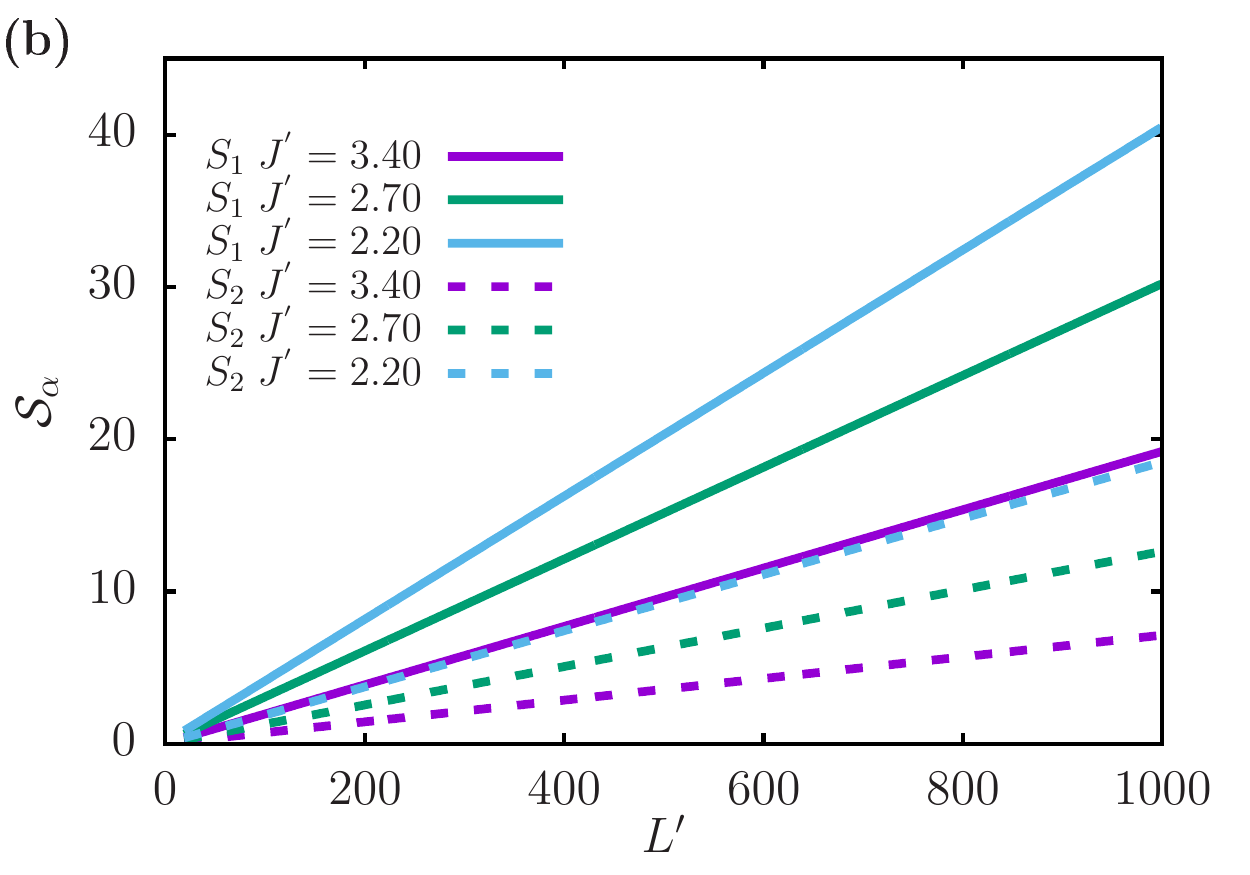}} 
\caption{(Color online) The von Neumann $\mathcal{S}_1$ (solid lines) 
  and second R\'enyi $\mathcal{S}_2$ (dashed lines) entanglement
  entropies as a function of the (line) subsystem size $L'$
  [Fig.~\ref{figlattice}(c)] for the VBS ground states of the
  (a) columnar-dimer and (b) staggered-dimer models.
  Results for three different values of the
   exchange coupling $J'$ are shown: 
   $J'= 2.20$ (blue),
   $J'= 2.70$ (green), and
   $J'= 3.40$ (magenta).} 
\label{figRenyi}
\end{figure}

In principle, the eigenvalues $\mu^2_m$ of the correlation matrix
$C$ are numerically calculated, see, e.g., Refs.~\cite{song11,alet15}. 
However, for a one-dimensional (line) subsystem $A$, 
the eigenvalues of the correlation matrix can be analytically
determined \cite{luitz15,alet15} since the correlation matrix $C$
is indeed a circulant
matrix \cite{gray}: In this case, the eigenvalues $\mu^2_m$ 
are given by the Fourier transform of the first line of the
correlation matrix $C$. For both dimer models, one finds
(see appendix \ref{ap:linesubsystem} for details)
\begin{equation}
    \mu_{m}^2 =    
       \left( \frac{1}{N_y} \sum_{k_y} \frac{A(m,k_y)}{\Omega(m,k_y)} \right)^2  
    - \left( \frac{1}{N_y} \sum_{k_y} \frac{B(m,k_y)}{\Omega(m,k_y)} \right)^2,                
\label{muAB}
\end{equation}
where the index $m = 1, 2, \cdots, N_A$ 
is related to the momentum $k_x$ parallel to the
system-subsystem boundary, 
\begin{equation}   
   k_x = -\frac{\pi}{2} + \frac{2\pi (m - 1)}{L'+2}, 
\label{kx-par}
\end{equation}
with $N' = N_AN_y$ and $N_A = (L' + 2)/2$.
Therefore, for a line subsystem $A$, the $N_A$ eigenvalues $\mu^2_m$ of the
correlation matrix $C$ can be easily expressed in terms of the coefficients 
$A_{\bk}$ and $B_{\bk}$ of the effective boson model \eqref{h-harm}. 
Once the sum over the momentum component $k_y$ is evaluated by
changing it to an integral, the entanglement entropies follow from
Eqs.~\eqref{Smu-neumann} and \eqref{Sqmu}.

\begin{table}[b]
\centering
\caption{Coefficients $a$, $b$, and $c$ obtained by fitting the von
  Neumann entanglement entropies $\mathcal{S}_1$ shown in
  Figs.~\ref{figRenyi}(a) and (b) with the curve \eqref{fit-ee}.} 
\begin{tabular}{lcccccccc}
\hline\hline
 &&  & Columnar  &  && &  Staggered & \\ 
$J'$ && a & b & c && a & b & c \\ 
\hline 
   3.40 &\quad \quad\quad\quad &  0.03  &  7.31e-09  &  0.04 
           &\quad\quad\quad&             0.02  &  1.16e-05  &  0.04 \\ 

  2.70 &  & 0.04  & 6.18e-09 & 0.07  && 0.03 & 1.07e-05 & 0.06\\ 

  2.20 &  & 0.06  & 3.78e-08 & 0.09  && 0.04 & 5.84e-05 & 0.08 \\ 
\hline\hline
\end{tabular} 

\label{tb-coefs-c}
\end{table}

\begin{table}[b]
\centering
\caption{Coefficients $a$, $b$, and $c$ obtained by fitting the second
  R\'enyi entanglement entropies $\mathcal{S}_2$ shown in
  Figs.~\ref{figRenyi}(a) and (b) with the curve \eqref{fit-ee}.} 
\begin{tabular}{lcccccccc}
\hline\hline
 &&  & Columnar  &  && &  Staggered & \\ 
$J'$ && a & b & c && a & b & c \\ 
\hline 
   3.40 &\quad \quad\quad\quad &  0.01  &  3.03e-09  &  0.02 
           &\quad\quad\quad&             0.01  &    5.82e-06  &  0.01 \\ 

   2.70 &  & 0.02  & 8.62e-10 & 0.03  && 0.01 & 6.44e-06 & 0.02\\ 

   2.20 &  & 0.03  & 1.39e-08 & 0.04  && 0.02 & 5.08e-08 & 0.04 \\ 
\hline\hline
\end{tabular} 
\label{tb-coefs-s}
\end{table}

From Eqs.~\eqref{muAB} and \eqref{kx-par}, one clearly sees how the 
finite-size nature of the subsystem $A$ is included in the calculation
of the entanglement entropy within our approach.
Recall that the coefficients $A_{\bk}$ and $B_{\bk}$ are obtained
within the bond-operator method, a formalism suitable to describe
the VBS phases in the thermodynamic limit. 
Similar considerations hold for Refs.~\cite{song11,luitz15,alet15}, 
where the entanglement entropies for magnetic ordered phases are
calculated via a modified spin--wave theory. 
One should note that in Refs.~\cite{song11,luitz15,alet15}, however,
an additional information about the subsystem size is encoded in a
staggered magnetic 
field $h$ that is added to restore the spin rotational symmetry of the
finite-size subsystem $A$. The value of $h$ is determined by
imposing that, at each site $i$ of the lattice, the $z$-component of the
spin operator $\langle S^z_i \rangle = 0$ [see also Sec.~VII from
Ref.~\cite{frerot15} for an alternative procedure].  
As discussed in Ref.~\cite{alet15}, this staggered magnetic field $h$
is an important ingredient to find the prefactor of the logarithmic
correction to the area law proportional to the number of Goldstone
modes [second term of Eq.~\eqref{s-neel}].  
Here, for the VBS phases, such an additional magnetic field is not
necessary, since these phases preserve the spin rotational symmetry.

Figure \ref{figRenyi} shows the von Neumann $\mathcal{S}_1$ 
and second R\'enyi $\mathcal{S}_2$ bipartite entanglement entropies in
terms of the subsystem size $L'$ for the VBS ground states of the
columnar-dimer [Fig.~\ref{figRenyi}(a)]  
and the staggered-dimer [Fig.~\ref{figRenyi}(b)]  models. 
We consider one-dimensional subsystems $A$ with sizes up to 
$L' = 1000$ and show the results for  
three different values of the exchange coupling $J'$.
Notice that $\mathcal{S}_1$ is larger than $\mathcal{S}_2$ for the same
value of $J'$, a feature that has been found for the magnetic ordered
phase of Heisenberg AFMs \cite{song11,luitz15,alet15}.   
Moreover, both entropies $\mathcal{S}_1$ and $\mathcal{S}_2$ are
dominated by an area law as expected for two-dimensional gapped phases
\cite{rmp-area-law,wen-book}.   
Indeed, we fit the data shown in Figs.~\ref{figRenyi}(a) and (b)  with the curve 
\begin{equation}
  S_\alpha = a L' + b\ln L' + c
\label{fit-ee}
\end{equation}
and, for the three values of the exchange coupling $J'$, we find that
$b < 10^{-5}$, see tables \ref{tb-coefs-c} and \ref{tb-coefs-s} for details. 
Finally, one sees that the prefactor $b$ of the logarithmic term is smaller
for the columnar-dimer model than for the staggered-dimer one
for both entanglement entropies $\mathcal{S}_1$ and $\mathcal{S}_2$.

\begin{figure}[t]
\centerline{\includegraphics[width=7.5cm]{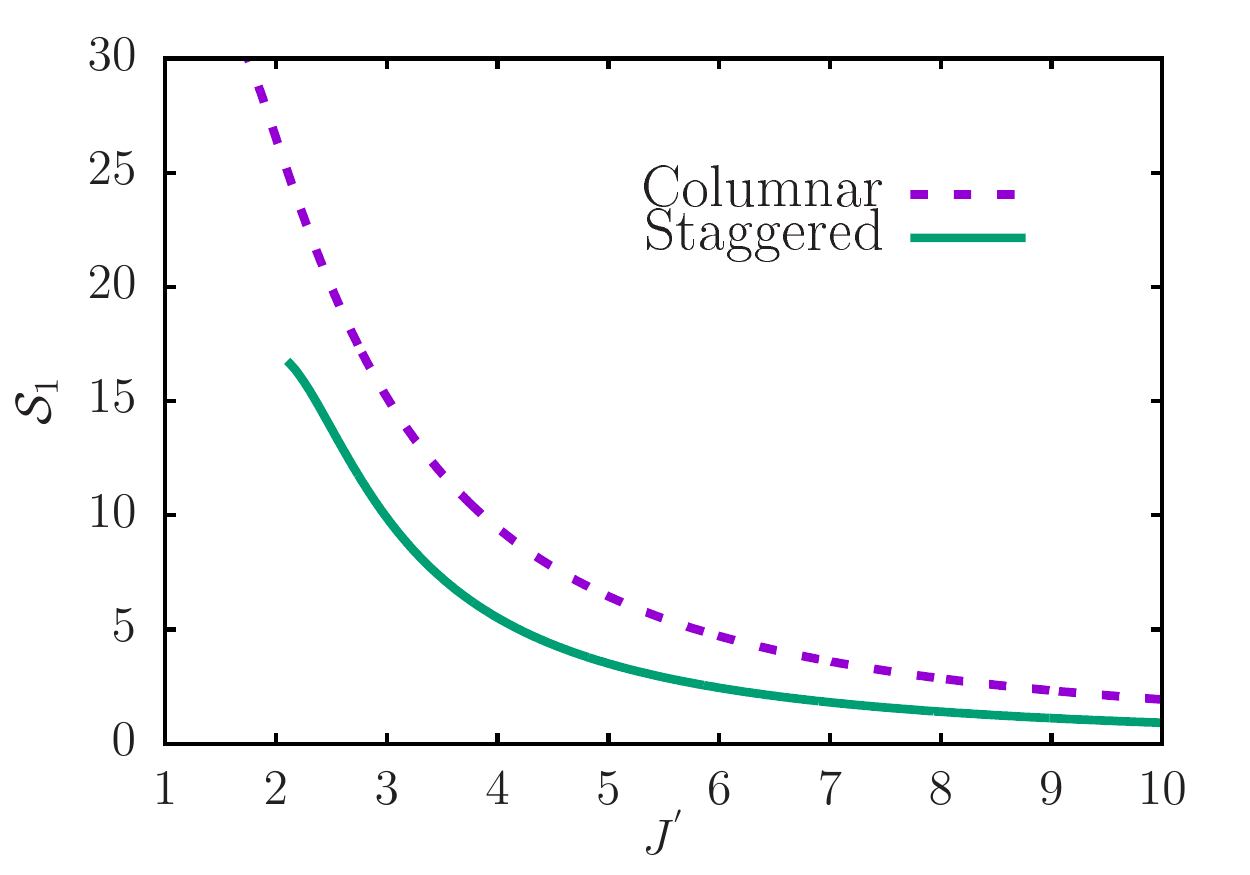}} 
\caption{(Color online) The von Neumman entanglement entropy
  $\mathcal{S}_1$ in terms of the exchange coupling $J'$ for the
  columnar-dimer (dashed magenta line) and the staggered-dimer (solid
  green line) models. Data for a line subsystem of size $L' = 400$.} 
\label{figS1}
\end{figure}

Although both bipartite entanglement entropies $\mathcal{S}_1$ and
$\mathcal{S}_2$ increase as $J'$ decreases, it seems that they do not
diverge as $J'$ approaches the N\'eel--VBS  quantum critical point. 
We illustrated such a behaviour in Fig.~\ref{figS1}, where 
it is shown the von Neumann entanglement entropy $\mathcal{S}_1$ 
as a function of the exchange coupling $J'$ for a subsystem $A$
with size $L' = 400$.
For both columnar-dimer and staggered-dimer models, one sees that
$\mathcal{S}_1$ has the same qualitatively behaviour, although its is
larger for the columnar-dimer model than for the staggered-dimer
one. 
For both dimer models, $\mathcal{S}_1$ 
reaches a maximum value at the smallest exchange coupling $J'$
determined via the numerical solutions of the self-consistent
equations \eqref{autoconsistencia}
[$J' = 1.70$ (columnar-dimer) and $J' = 2.00$ (staggered-dimer), see
Sec.~\ref{sec:self}], 
a feature that indicates a possible absence of divergence at the
quantum critical point.
One should mention that such an absence of divergence of the
entanglement entropy at criticality was previously observed in the
N\'eel--VBS QPT of a two-dimensional bilayer Heisenberg AFM
\cite{wessel14} and in the superfluid-Mott insulator QPT of a
two-dimensional Bose-Hubbard model \cite{frerot16}.

Finally, one also sees in Fig.~\ref{figS1} that, for larger values
of $J'$, the entanglement entropy 
$\mathcal{S}_1$ slowly decreases. Indeed, one finds, e.g., for the
columnar-dimer model, that $\mathcal{S}_1 = 0.57$, $0.27$, and $0.16$
respectively for $J' = 20.0$, $30.0$, and $40.0$. Such a behaviour is
expected since, as the (intra-dimer) exchange coupling $J'$ increases,
the dimers get more and more isolated and, therefore, $\mathcal{S}_1$
should vanish in this limit.

\subsection{Entanglement spectra for the VBS phases}
\label{sec:es-vbs}

In addition to the bipartite entanglement entropies $\mathcal{S}_1$
and $\mathcal{S}_2$, the procedure employed in our work allow us to easily
calculate the entanglement spectrum $\epsilon_{k_x}$ as defined in
Eq.~\eqref{ent-h}. Notice that once the eigenvalues $\mu^2_m =
\mu^2_{k_x}$ of the correlation matrix $C$ are known, 
the entanglement spectrum $\epsilon_{k_x}$ follows from
Eq.~\eqref{ekck}.

\begin{figure}[t]
\centerline{\includegraphics[width=7.5cm]{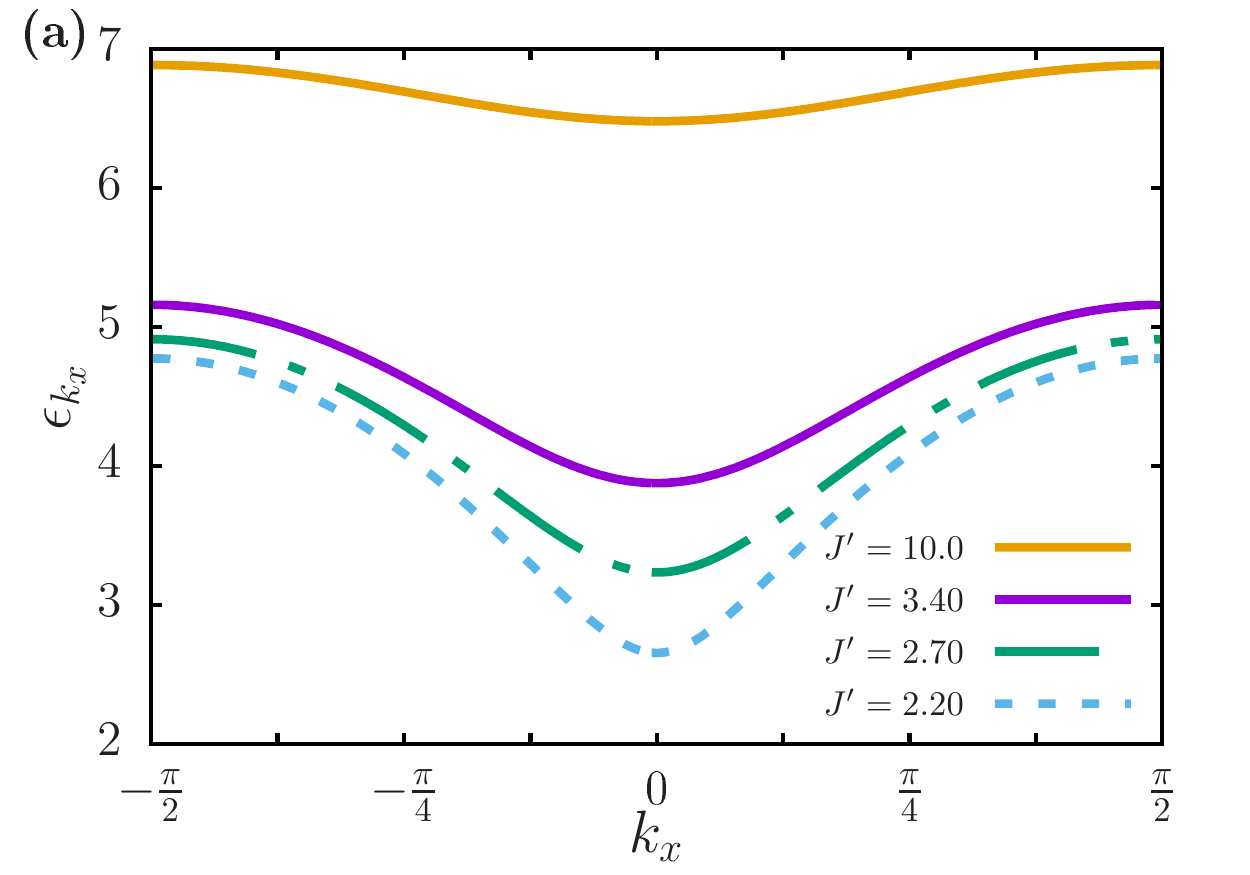}} 
\vskip0.4cm
\centerline{\includegraphics[width=7.5cm]{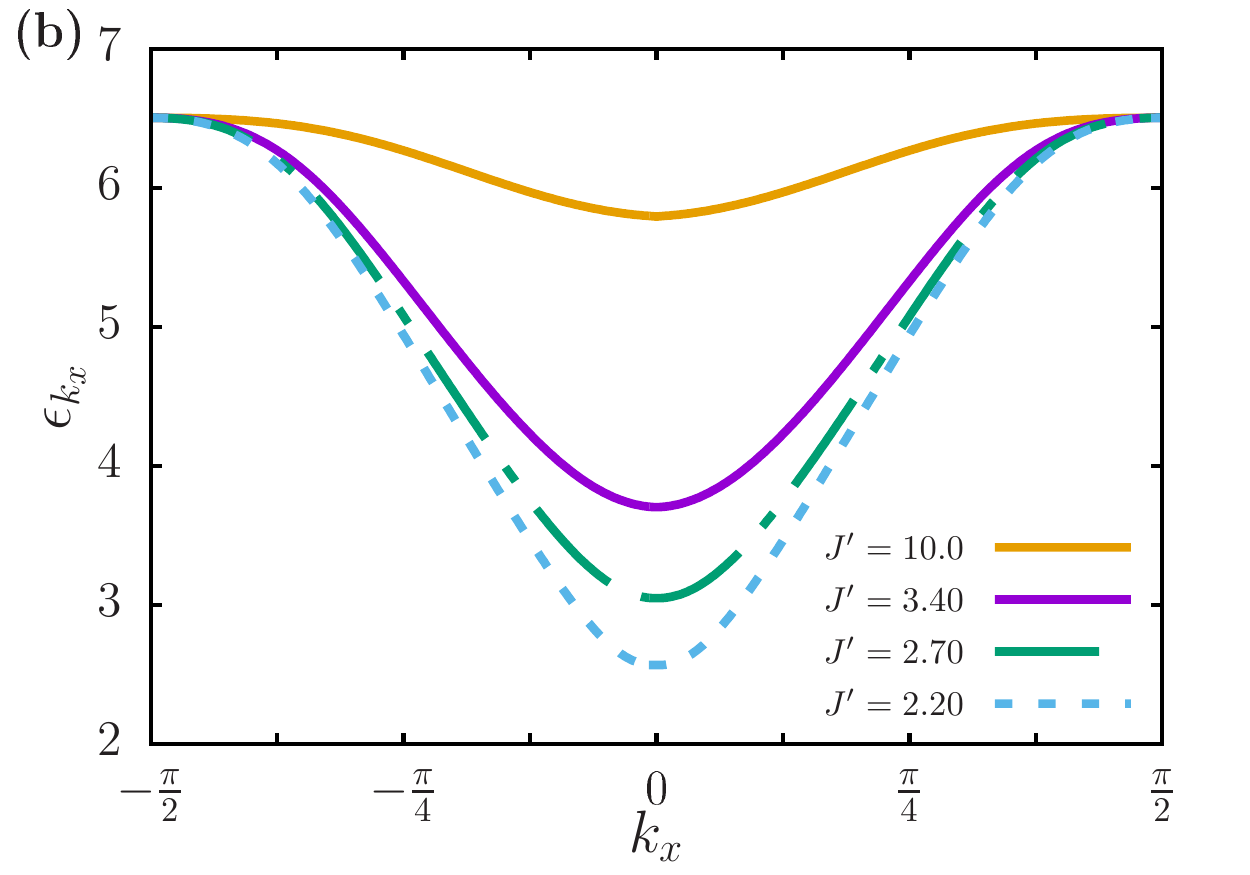}} 
\caption{ (Color online) Entanglement spectrum $\epsilon_k$
  [Eq.~\eqref{ent-h}] in terms of the momentum $k_x$ parallel to the
  system-subsystem boundary \eqref{kx-par} of the (a) columnar-dimer
  and (b) staggered-dimer models. 
  Data for a line subsystem of size $L' = 1000$.
  Results for four different values of the exchange coupling $J'$ are
  displayed: 
  $J'= 2.20$ (dashed blue line),
  $J'= 2.70$ (dotted-dashed green line), 
  $J'= 3.40$ (solid magenta line), and
  $J'=10.0$ (solid orange line).} 
\label{fig-es}
\end{figure}

In Fig.~\ref{fig-es}, we show the entanglement spectra of the
columnar-dimer [Fig.~\ref{fig-es}(a)] and staggered-dimer [Fig.~\ref{fig-es}(b)]
models for four different values of the exchange coupling $J'$ 
(a line subsystem $A$ with size $L' = 1000$ is considered).
One sees that the entanglement spectra of the two dimer models are
qualitatively similar, although the bandwidth is larger for the
staggered-dimer model than for the columnar-dimer one at the same value of
the exchange coupling $J'$.  
For both dimer models, the energy gap of the
entanglement spectra is at $k_x=0$ and the gap decreases as $J'$
approaches the critical coupling $J'_c$. Differently from the triplon
spectrum $\Omega_\bk$, the gap of the entanglement spectrum does not 
close as the system approaches the N\'eel--VBS quantum critical point: 
we find
$\epsilon_{k_x=0} = 1.77$ ($J' = 1.70$, columnar-dimer) and
$\epsilon_{k_x=0} = 2.47$ ($J' = 2.00$, staggered-dimer),
compare with Fig.~\ref{fig:gaps}.   
Such a feature is in contrast with the behaviour of the
two-dimensional Bose-Hubbard model which also displays a QPT between
a gapped (Mott-insulator) and a gapless (superfluid) phases \cite{frerot16}:
it was found that the gap of the entanglement spectrum closes at the
superfluid--Mott insulator QPT driven by the ratio $t/U$ between the
tunneling amplitude $t$ and the on-site repulsion $U$ at integer
(fixed) filling.
Finally, for larger values of the exchange coupling $J'$, the
entanglement spectrum is almost flat. 
We exemplify this feature in Fig.~\ref{fig-es}, where the
entanglement spectra for both dimer models with exchange
coupling $J'=10.0$ are shown.  
This behaviour is indeed in agreement with the fact that the
dimers are almost isolated for $J' \gg 1$, see discussion at the end
of Sec.~\ref{sec:ee-vbs}.

After the proposal of Li and Haldane \cite{haldane-es} that the
low-lying entanglement spectrum can be used to identify topological
order, a series of papers has been devoted to study the entanglement
properties of two-dimensional topological phases, see, e.g., Ref.~[2]
from Ref.~\cite{alba13}. On the other hand, the more conventional
phases realized in two-dimensional systems have been received less
attention \cite{review-nicolas}. In the latter case, an interesting result is due to
Alba {\it et al.} \cite{alba13} who showed that the entanglement
spectrum of the Mott-insulator phase of the square lattice two-dimensional
Bose-Hubbard model is dominated by degrees of freedom located at the  
system-subsystem boundary. In particular, the entanglement spectrum
can be interpreted as the spectrum of a (boundary)
tight-binding model whose sites are at the system-subsystem boundary.  
Assuming that this is indeed a quite general feature of a gapped phase,
we expect that the entanglement spectra shown in Figs.~\ref{fig-es}(a)
and (b), which are derived for line subsystems $A$, might be
characteristic of the columnar-dimer and staggered-dimer models, i.e.,
such a features might be found in entanglement studies regardless the  
subsystem shape.

\section{Summary}
\label{sec:summary}

In this paper, we have studied the (quantum paramagnet) VBS phases of
the columnar-dimer and staggered-dimer Heisenberg AFMs on a square
lattice within the bond-operator formalism at the harmonic
approximation. In particular, these results, combined with a procedure employed in
Refs.~\cite{song11,luitz15,alet15} for magnetic ordered phases, allowed
us to calculate the bipartite von Neumann and second R\'enyi
entanglement entropies for the VBS ground states of the two dimer
models. Choosing an one-dimensional (line) subsystem $A$, this
formalism provides the area law behaviour 
for the entanglement entropies as expected for gapped phases.

It would be interesting to apply the bond-operator based approach
discussed here, e.g., for rectangular strip and square subsystems $A$
[see, e.g., Fig.~1 from Ref.~\cite{alet15}]. Such studies would allow us
to check whether our results for the entanglement entropies depend on
the shape of the subsystem $A$. In this case, however, the eigenvalues
$\mu^2_k$ of the correlation matrix $C$ should be numerically determined.
Moreover, it would be important to determine the effects of the cubic
$H_3$ [Eq.~\eqref{hk3}] and quartic $H_4$ [Eq.~\eqref{hk4}] terms of the
effective boson model \eqref{eff-h} on the entanglement
entropies. These two terms could be perturbatively considered as done,
e.g., in Ref.~\cite{doretto12}. Recall the role of the cubic term in
the distinction between the two dimer models as discussed in
Sec.~\ref{sec:model}. As expected, the mean-field results obtained
here are qualitatively similar for both dimer models.  
We intend to performed these two studies in a future publication.

Finally, it would also be interesting to compare the entanglement
spectra derived here with the ones determined via density matrix
renormalization group (DMRG) calculations as done, e.g., for the
square lattice Bose--Hubbard model \cite{alba13}.
However, as far as we know, such DMRG data for the columnar-dimer and
staggered-dimer models are not available at the moment.

\acknowledgments

We thank E. Miranda for helpful discussions.
L.S.G.L. kindly acknowledges the financial support of the
Coordena\c{c}\~ao de Aperfei\c{c}oamento 
de Pessoal de N\'ivel Superior - Brasil (CAPES) - Finance Code 001.

\appendix

\section{Classical dimerized Heisenberg models}
\label{ap:classical}

The classical phases of the columnar-dimer
[Eq.~\eqref{h1}] and staggered-dimer [Eq.~\eqref{h2}] models can be
determined by parametrizing the spins $\bS^1_i$ and $\bS^2_i$ [see
Figs.~\ref{figlattice}(a) and (b)] as
\begin{align}
  \bS_i^1 = \hat{e}_1 \cos(\bQ \cdot \bR_i) + \hat{e}_2 \sin(\bQ \cdot\bR_i), 
\nonumber \\
\label{s1s2} \\
  \bS_i^2 = \hat{e}_3 \cos(\bQ \cdot \bR_i) + \hat{e}_4 \sin(\bQ \cdot\bR_i) ,
\nonumber 
\end{align}	
where $\bQ$ is the ordering wave vector, $\bR_i$ is a vector of
the dimerized lattice, and the set of unit vectors  $\hat{e}_i$ obeys
the following relations 
\begin{eqnarray}
  \hat{e}_1 \cdot  \hat{e}_2 &=& \hat{e}_3 \cdot \hat{e}_4 = 0,
\nonumber \\
  \hat{e}_1 \cdot  \hat{e}_3 &=& \hat{e}_2 \cdot \hat{e}_4 = \cos\theta ,  
  \quad \hat{e}_2 \cdot \hat{e}_3 = - \hat{e}_1 \cdot \hat{e}_4 = \sin \theta.
\nonumber
\end{eqnarray}

Substituting Eq.~\eqref{s1s2} into the Hamiltonian \eqref{h1} of the
columnar-dimer model, we obtain the energy $E$ as a function of the
components $Q_x$ and $Q_y$ of the ordering wave vector and the
angle $\theta$, namely
\begin{align}
   E = &  J' \frac{N}{2} \cos \theta  +\frac{N}{2} [ 2 \cos(Q_y)   \nonumber \\
          & + \cos \theta \cos(2 Q_x) + \sin \theta \sin(2 Q_x) ].
\label{classical-e}
\end{align}
The ground state energy follows from the minimization of
Eq.~\eqref{classical-e} with respect to the parameters $Q_x$, $Q_y$,
and $\theta$. For $J' > 0$, we find $\bQ = (0,\pi)$ and $\theta =
\pi$, i.e., 
\begin{equation}
	\bS_i^1 = \hat{e}_1\cos(\pi R_{y,i})
        \quad  {\rm and} \quad \bS_i^2 = \hat{e}_3\cos(\pi R_{y,i}).
\label{conf-col}
\end{equation}
With the aid of Fig.~\ref{figlattice}(a), one easily sees that the
configuration \eqref{conf-col} corresponds to a collinear N\'eel phase.

Similarly, substituting Eq.~\eqref{s1s2} into the Hamiltonian
\eqref{h2} of the staggered-dimer model, we arrive at  
\begin{align}
  E = & J' \frac{N}{2} \cos \theta  + \frac{N}{2} [  \cos\theta \sum_\tau \cos(\bQ \cdot \taub)  
\nonumber \\
	&+ \sin \theta \sum_\tau \sin(\bQ \cdot \taub) ],
\end{align}
where the dimer nearest-neighbor vectors $\taub$ are given by
Eq.~\eqref{tau-stag}. In this case, we find that $\bQ =  (0,0)$ and
$\theta = \pi$, i.e., 
\begin{equation}
  \bS_i^1 = \hat{e}_1, \quad {\rm and} \quad \bS_i^2 = \hat{e}_3.
\end{equation}

\section{Entanglement spectrum and the correlation matrix}
\label{ap:pseudoenergy}

In this section, we derive the relation \eqref{ekck} between the eigenvalues
$\mu^2_k$ of the correlation matrix $C$ and the entanglement spectrum
$\epsilon_k$.  

As mentioned in Sec.~\ref{sec:ee}, the generic quadratic Hamiltonian
\eqref{generalH1} and the corresponding entanglement Hamiltonian
\eqref{generalH2} can be diagonalized by the same Bogoliubov
transformation. Let us consider
\begin{align}
   b_k = &\frac{1}{2} \sum_i (\phi_{ki} + \psi_{ki} ) a_i + (\phi_{ki} - \psi_{ki} )  a_i^\dagger, 
\nonumber \\
   b_k^\dagger = &\frac{1}{2} \sum_i (\phi_{ki} + \psi_{ki} ) a_i^\dagger + (\phi_{ki} - \psi_{ki} )  a_i, 
\label{Atransformation}
\end{align} 
where the coefficients $\phi_{ki}$ and $\psi_{ki}$ are assumed to be real
for simplicity. Due to the bosonic algebra of the $b_k$ operators, the
coefficients $\phi_{ki}$ and $\psi_{ki}$ obey the relations 
\begin{align}
  & [ b_k, b_p^\dagger] = \frac{1}{2} \sum_i \phi_{ki} \psi_{pi} + \psi_{ki} \phi_{pi} = \delta_{kp},
\nonumber \\
  & [ b_k, b_p ] = \frac{1}{2} \sum_i \psi_{ki} \phi_{pi} - \phi_{ki} \psi_{pi} = 0,
\nonumber
\end{align}
which imply that 
\begin{equation}
 \sum_i \phi_{ki} \psi_{pi}  = \sum_i \psi_{ki} \phi_{pi} = \delta_{kp}.
\label{Aortogonal}  
\end{equation}

For sites $i$ and $j$ associated with the subsystem $A$, we have
\begin{eqnarray}
 \langle a^\dagger_i a_j \rangle &=& {\rm Tr}\left( \rho \; a^\dagger_i a_j \right)
\nonumber \\
       &=& \sum_A \langle \xi_A| 
                \left( \sum_{\bar{A}} \langle \xi_{\bar{A}} | \rho | \xi_{\bar{A}} \rangle  \right)  a^\dagger_i a_j | \xi_A \rangle         
\nonumber \\
      &=& {\rm Tr}_A  \left( \rho_A \; a^\dagger_i a_j \right),
\label{average-a}
\end{eqnarray}
where the states $|\xi_A\rangle$ and $|\xi_{\bar{A}}\rangle$ are
respectively associated with the subsystem $A$ and its complementary
$\bar{A}$ as defined in Sec.~\ref{sec:intro}.
Moreover, since the Bogoliubov transformation \eqref{Atransformation}
diagonalizes the entanglement Hamiltonian $\mathcal{H}_E$
[Eq.~\eqref{ent-h}] and the reduced density matriz $\rho_A$ has the
form \eqref{rhoDiagonal}, we have 
\begin{align}
  {\rm Tr}_A \left( \rho_A b_k^\dagger b_q \right) &=\frac{1}{e^{\epsilon_k} -1 } \: \delta_{kq} ,
\nonumber \\
  {\rm Tr}_A \left( \rho_A b_k^\dagger b_q^\dagger \right) &= 0.
\label{average-b}
\end{align}
Therefore, from Eqs.~\eqref{average-a} and \eqref{average-b} and the inverse
of the transformation \eqref{Atransformation}, one shows that 
\begin{equation}
 \langle a_i^{\dagger} a_j\rangle + \frac{\delta_{ij}}{2}  =
        \frac{1}{4} \sum_k (\phi_{ki} \phi_{kj} + \psi_{ki} \psi_{kj}) \: \coth \left( \frac{\epsilon_k}{2} \right)
\label{Acij}
\end{equation}
and, similarly,  
\begin{align}
 \langle a_i^{\dagger} a_j^{\dagger}\rangle &= \langle a_i a_j\rangle 
\nonumber \\
      &= - \frac{1}{4} \sum_k (\phi_{ki} \phi_{kj} - \psi_{ki} \psi_{kj}) \: \coth \left( \frac{\epsilon_k}{2} \right).
\label{Afij}
\end{align}

The single-particle Green's functions \eqref{greenf} assume the form 
\begin{align}
    f_{ij} + g_{ij}  = \frac{1}{2} \sum_k \psi_{ki} \psi_{kj} \coth \left( \frac{\epsilon_k}{2} \right),	
\nonumber  \\
    f_{ij} - g_{ij} = \frac{1}{2} \sum_k \phi_{ki} \phi_{kj} \coth \left( \frac{\epsilon_k}{2} \right), 
\end{align}
where $i$ and $j$ $\in \, A$. With the aid of the orthogonality
condition \eqref{Aortogonal}, one shows that 
\begin{align}
  \sum_i 2 \left( f_{ij} + g_{ij} \right) \phi_{ki} = \coth\left( \frac{\epsilon_k}{2} \right) \psi_{kj}
       \equiv \mu_k \psi_{kj},
\nonumber \\
  \sum_i  2 \left( f_{ij}  - g_{ij} \right) \psi_{ki} = \coth\left( \frac{\epsilon_k}{2} \right) \phi_{kj}
       \equiv \mu_k \phi_{kj}. 
\end{align}
The above equation can be written in a matrix form
\begin{align}
	&\hat{\phi}_{k} \: G^{++}  = \mu_k \hat{\psi}_{k},
\nonumber \\
	&\hat{\psi}_{k} \: G^{--}  = \mu_k  \hat{\phi}_{k},
\label{eigen-g}
\end{align}
where the elements of the $N_A \times N_A$ matrices $G^{++}$ and $G^{--}$
are given by   
\begin{align}
  &G^{++}_{ij} = + \langle ( a_i^\dagger + a_i )( a_j^\dagger + a_j )\rangle
                    = 2f_{ij} + 2g_{ij},
\nonumber \\
  &G^{--}_{ij} = - \langle ( a_i^\dagger - a_i )( a_j^\dagger - a_j ) \rangle 
                   = 2f_{ij} - 2g_{ij},
\nonumber
\end{align}
and the vectors $\hat{\psi}_{k}$ and $\hat{\phi}_{k}$ are defined as
\begin{align}
\hat{\psi}^t_k = \left( \psi_{k1} \; \psi_{k2} \; \cdots \psi_{kN_A} \right),
\nonumber \\
\hat{\phi}^t_k = \left( \phi_{k1} \; \phi_{k2} \; \cdots \phi_{kN_A} \right).
\nonumber
\end{align}
From Eq.~\eqref{eigen-g}, we find the eigenvalue equation for the
correlation matrix C, 
\begin{equation}
   \hat{\phi}_k G^{++} G^{--} =  \hat{\phi}_k C =\mu_k^2 \hat{\phi}_k,
\label{G++}
\end{equation}
that provides the relation \eqref{ekck} between the eigenvalues
$\mu^2_k$ of the correlation matrix $C$ and the entanglement spectrum
$\epsilon_k$.  

Finally, one notices that 
\begin{align}
   & C_{ij} = [G^{++}G^{--}]_{ij} = 4\sum_{s \in A}  \left( f_{is} + g_{is} \right) \left( f_{sj} - g_{sj} \right) 
\nonumber \\
   & = \sum_{k,p}\psi_{ki} \left( \sum_s  \psi_{ks} \phi_{ps} \right) \phi_{pj}  
         \coth \left( \frac{\epsilon_k}{2} \right) \coth \left( \frac{\epsilon_p}{2} \right) 
\nonumber \\
   & = \sum_k \psi_{ki} \phi_{kj} \coth^2 \left( \frac{\epsilon_k}{2} \right),
\end{align}
which is the bosonic version of Eq.~(16) from Ref.~\cite{peschel03}
written in a slightly different notation.

\section{Eigenvalues of the correlation matrix for one-dimensional (line) subsystem}
\label{ap:linesubsystem}

For an arbitrary subsystem $A$, the correlation matrix
\eqref{correlationMatrix} satisfies the property  
$C_{ij} = C_{|i-j|}$, as one can easily see from Eq.~\eqref{transf-f-g}, 
indicating that the correlation matrix $C$ is a 
Toeplitz matrix \cite{gray}. In particular, for an one-dimensional
subsystem $A$ with periodic boundary conditions, one finds that the
correlation matrix $C$ is indeed a circulant matrix: 
in this case, each row of the matrix can be obtained from the first
row by a shift of the matrix elements \cite{gray}. 
Due to this translational property, the eigenvalues of a circulant
matrix can be obtained by a discrete Fourier transform of its first
row elements.

Let us consider the one-dimensional subsystem $A$ shown in
Fig.~\ref{figlattice}(c), i.e., a chain of size $L'$ and 
$N_A = (L'+ 2)/2$ sites. Since the vectors of the underline dimerized
lattice are $\bR_i = 2i\hat{x}$, with $i = 1,2, \ldots, N_A$, 
the expression \eqref{transf-f-g} for the single-particle Green's
function $f_{ij} $ and $g_{ij}$ can be written as 
\begin{align}
  &f_{ij} = +\frac{1}{2N_A} \sum_{k_x}  \alpha_{k_x}  \cos[ 2k_x (i - j)], 
\nonumber \\
  &g_{ij} = -\frac{1}{2N_A} \sum_{k_x}  \beta_{k_x} \cos[ 2k_x (i - j)],
\label{transf-f-g-2}
\end{align}
where the functions $\alpha_{k_x}$ and $\beta_{k_x}$ are defined as 
\begin{equation}
  \alpha_{k_x} = \frac{1}{N_y}\sum_{k_y} \frac{A_\bk}{\Omega_\bk} 
  \quad {\rm and} \quad 
 \beta_{k_x} = \frac{1}{N_y}\sum_{k_y} \frac{B_\bk}{\Omega_\bk}.
\end{equation}

The eigenvalues $\mu^2_k$ of the correlation matrix C are given by the  
discrete Fourier transform \cite{gray}
\begin{equation}
   \mu_m^2 = \sum_{j=0}^{N_{A}-1} C_{0j} \:  e^{ - 2 \pi i jm/N_{A}},  
\label{eigenC01}
\end{equation}
where $C_{0j}$ are the elements of the first row of the correlation matrix
\eqref{correlationMatrix}. Using the convolution property $(f*g)(x) =
(g*f)(x)$, one shows that the matrix elements \eqref{correlationMatrix}
can be written as  
\begin{equation}
   C_{0j} = C(l) = 4 \sum_{x=0}^{N_A - 1} f_x  \: f_{l-x} - g_x  \: g_{l-x},
\label{first-row}
\end{equation}
where $s - i = x$, $j - i = l$, and $j - s = l-x$.
Therefore, Eq.~\eqref{eigenC01} assumes the form
\begin{equation}
  \mu_m^2 = \sum_{j=0}^{N_{A}-1} C(l) \:  e^{ - 2 \pi i lm/N_{A}},
\label{eigenC02}
\end{equation}
Finally, from Eqs.~\eqref{transf-f-g-2}, \eqref{first-row}, and \eqref{eigenC02}  
and with the aid of the Fourier transform property 
$\mathcal{F}\{(f*f)(x)\}_l = \mathcal{F}\{(f)(x)\}_l \mathcal{F}\{(f)(x)\}_l$, 
one shows that
\begin{equation}
    \mu_{m}^2 = \alpha_{k_x}^2 - \beta_{k_x}^2 
\end{equation}
with $k_x$ given by Eq.~\eqref{kx-par}.


\end{document}